# Of heading, posture and body rotations derived from data acquired by animal-borne accelerometers, magnetometers and gyrometers, kernel density estimation of the corresponding spherical distributions, and fine-scale movement reconstruction

Simon Benhamou

CEFE – CNRS Montpellier, France

Associated to Cogitamus Lab

simon.benhamou@cefe.cnrs.fr

**Abstract.** This paper provides the mathematical expressions, formulated in classical trigonometric terms, that are required (i) to compute an animal's body orientations and rotations in 3D space from data acquired at high-frequency by on-board accelerometers, magnetometers and gyrometers, (ii) to relate 3D rotations to changes in 3D orientation, (iii) to represent their spherical distributions through kernel density estimation (KDE), and (iv) to reconstruct fine-scale movements. The way the huge amount of data that is acquired by these on-board devices can be more easily managed is also considered.

## 1. Introduction

Thanks to a number of technological improvements, there has been an increasing interest in deploying animal-borne loggers to investigate how animals move at various scales, and thus in attempting to elucidate the underlying behaviors (Williams et al. 2020). In particular, three types of high-frequency (>10 Hz) tri-axial bio-logger are commonly used nowadays to study movements performed at very small scale: accelerometers, magnetometers and gyrometers, which make it possible to assess 3D orientations and rotations.

My aim here is to provide a consistent and rational guide to the treatment of the data acquired by these loggers. Orientations and rotations in 3D space, as tackled in aeronautics or robotics, are usually expressed in terms of mathematical formulae that are too complex (e.g. involving quaternions) to be of any real use to biologists/behavioral ecologists. Formulae involving classical trigonometric functions, more easily understandable, have been published in bio-logging oriented papers, but in a rather scattered way, and they are not always reliable. In what follows, I will provide the mathematical formulae, expressed in classical trigonometric terms, that are required (i) to compute an animal's body orientations and rotations in 3D space from data acquired at high-frequency by on-board accelerometers, magnetometers and gyrometers, (ii) to relate 3D rotations to changes in 3D orientation, (iii) to represent their spherical distributions through kernel density estimation (KDE), and (iv) to reconstruct fine-scale movements. I will also consider how to manage the huge amount of data that are acquired by these on-board devices. I will focus mainly on loggers assumed to have been tightly mounted on the animal (usually on its back). Much less information can be obtained when the logger is mounted on a collar, liable to roll around the neck. I will consider it only for reconstructing 2D fine-scale movements of a terrestrial animal.

Two frames of reference need to be considered. One is Earth-bound. It is defined by the three orthogonal axes *X*, *Y* and *Z*, which, throughout this paper, will conform the so-called East-North-Up (ENU) convention. The other is animal's body-bound. It is defined by the surge (*U*), sway (*V*) and heave (*W*) axes, corresponding respectively to the tail→head (posterio-anterior), right→left (transversal), and belly→back (ventro-dorsal) axes (Fig. 1). They are orthogonal to, respectively, the frontal (sway x heave) plane, the sagittal (surge x heave) plane and the coronal (surge x sway) plane (in the few species moving in a stand-up posture, such as humans, the frontal and coronal planes are swapped because the surge axis then corresponds to the back→belly rather than tail→head axis). The direction of the surge axis (tail→head) was chosen to reflect the usual forward movement direction. The directions of the sway (right→left) and heave axes (belly→back) were then defined in this way to have the relative orientations of the three body axes, *U*, *V* and *W*, that match those of the *X*, *Y* and *Z* axes, respectively. The three body directions therefore follow the 'right-hand rule' with the index, middle finger and thumb representing the surge, sway and heave axes respectively.

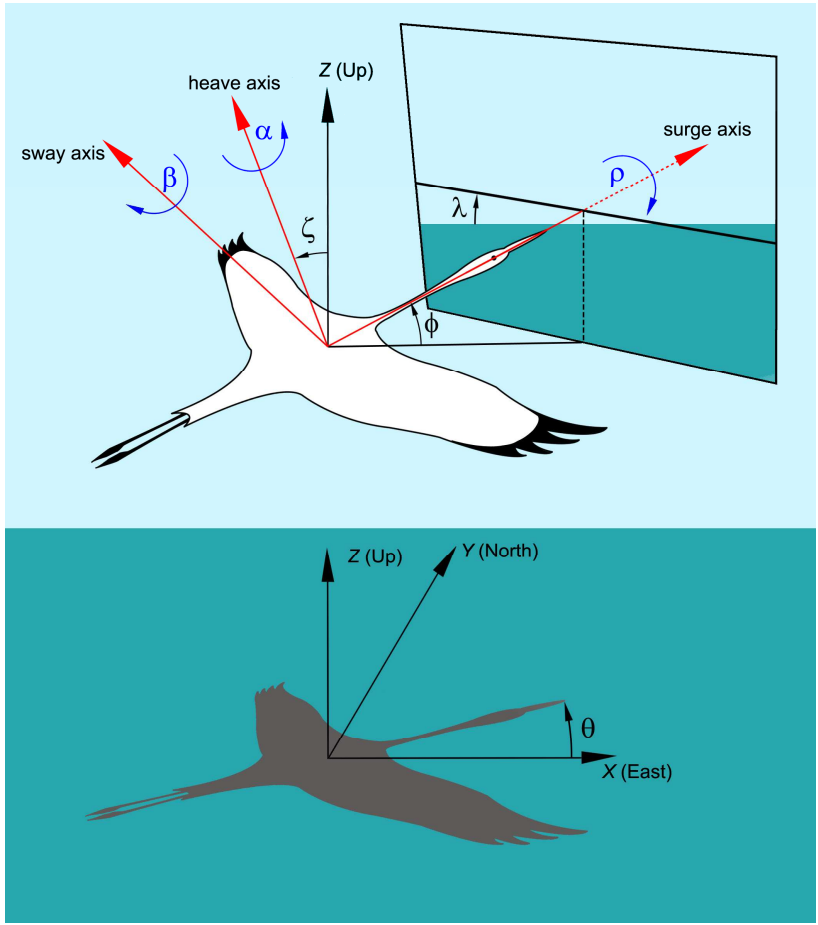

**Fig. 1. Representation of various axes and angles.** Let us consider a bird flying above a body of water. The *XY* plane, corresponding to the water surface, is horizontal, with the Earth-bound axes *X* pointing to East and *Y* pointing to North, while the Earth-bound axis Z points upwards (ENU convention). The animal-bound axes, referred to as the surge, sway and heave axes, correspond to the tail→head, right→left and belly→back axes, respectively. The 3D *orientation* is defined by the azimuth, θ, the elevation angle, φ, and the bank angle λ. The azimuth is the horizontal component of heading. It corresponds to the orientation of the surge axis in the *XY* plane. The elevation angle corresponds to the orientation of the surge axis, measured from the horizontal plane in the vertical plane at azimuth θ. It acts as both the vertical component of heading and the longitudinal component of posture. The bank angle is the transversal component of posture. It is measured from the horizontal level in the animal's frontal plane (i.e. orthogonally to the surge axis) and corresponds to the orientation of the sway axis in this plane. The overall body inclination, ζ, combines the longitudinal (φ) and transversal (λ) components of posture. It corresponds to the deviation of the heave axis from the *Z* axis. Yaw, α, pitch, β, and roll, ρ, are *rotations* done around the three main body axes.

Heading corresponds to the surge axis orientation expressed in the Earth-bound frame of reference. In 3D space, it is usually specified by two angular spherical coordinates, hereafter referred to as the azimuth, $\theta \in [-\pi, \pi]$, measured in the horizontal ($XY$) plane, and the elevation angle, $\phi \in [-\pi/2, \pi/2]$, measured vertically from the $XY$ plane. Classically, the terms 'azimuth' and 'elevation angle' correspond to the compass bearing and angular elevation above the horizon line of an actual object. By extension, they are used here to specify the heading as the compass bearing and angular elevation of a virtual object towards which the animal would face. The posture corresponds to the body orientation with respect to gravity. It is usually specified in terms of longitudinal and transversal inclinations. The former corresponds to the elevation angle, $\phi \in [-\pi/2, \pi/2]$, which is therefore a component of both heading and posture. The latter, $\lambda \in [-\pi, \pi]$, referred to as the bank angle, corresponds to the orientation of the sway axis with respect to the horizontal level in the frontal plane (i.e. orthogonally to the surge axis; Fig. 1). It is sometimes mistaken for the vertical orientation of the sway axis measured from the horizontal ($XY$) plane, $\phi' \in [-\pi/2, \pi/2]$ ($\sin(\lambda) = \sin(\phi')/\cos(\phi)$; see Eq. **3**b). Overall, the 3D orientation is therefore unequivocally specified by $\theta$, $\phi$, and $\lambda$. With the ENU convention, one gets $\theta = 0$ when the surge axis points eastwards and $\theta = \pi/2$ when it points northwards, and $\phi > 0$ and $\lambda > 0$ when the surge and sway axes, respectively, point above the horizon line. It is worth noting that, as Euler/Tait-Bryan angles, azimuth $\theta$, elevation angle $\phi$, and bank angle $\lambda$, which correspond to *orientations* expressed in the *Earth-bound* frame of reference, are involved in rotation matrices to link the current heading/posture to a 'neutral' one. This is probably why these orientations, or changes in them, are often referred to as yaw, pitch and roll whereas, to avoid any ambiguity, these latter terms should be reserved for *rotations* expressed in the *body-bound* frame of reference (see Fig.1 and §6).

In this paper, the key trigonometric relationships will be expressed using rotation matrices. They provide an easy way to determine the new Cartesian coordinates of a vector that was rotated by a given angle $\varphi$ in a fixed frame of reference or, equivalently, that was kept fixed while the frame of reference was rotated by $-\varphi$. Depending on whether the rotation is done around the first ($X$ or $U$), second ($Y$ or $V$) or third ($Z$ or $W$) axis, these matrices are:

$$\mathbf{R}_1(\varphi) = \begin{pmatrix} 1 & 0 & 0 \\ 0 & \cos(\varphi) & -\sin(\varphi) \\ 0 & \sin(\varphi) & \cos(\varphi) \end{pmatrix}, \ \mathbf{R}_2(\varphi) = \begin{pmatrix} \cos(\varphi) & 0 & -\sin(\varphi) \\ 0 & 1 & 0 \\ \sin(\varphi) & 0 & \cos(\varphi) \end{pmatrix}, \ \mathbf{R}_3(\varphi) = \begin{pmatrix} \cos(\varphi) & -\sin(\varphi) & 0 \\ \sin(\varphi) & \cos(\varphi) & 0 \\ 0 & 0 & 1 \end{pmatrix}$$

where, in agreement with the ENU convention, $\varphi$ is measured positively *counterclockwise* for $\mathbf{R}_1$ and $\mathbf{R}_3$ but *clockwise* for $\mathbf{R}_2$ when the rotation axis considered points towards the observer.

These matrices are involved in determining the vector $\mathbf{B} = (B_U, B_V, B_W)$ that applies to the animal's body from the vector $\mathbf{D} = (D_U, D_V, D_W)$ recorded by the device (corresponding to acceleration, geomagnetic field or angular velocity, depending on the type of device considered). One gets $\mathbf{B} = \mathbf{D}$ only if the surge, sway and heave axes of the body and of the device match each other. In what follows, the device will be assumed to be tightly mounted on the animal's back, unless otherwise specified. However, as the animal's back may not be flat, the device may have been mounted somewhat tilted longitudinally (around the sway axis) by $\phi^*$, and/or transversally (around the surge axis) by $\lambda^*$. A mismatch may also occur in terms of azimuth. The general procedure required to derive **B** from **D** is quite challenging (Johnson & Tyack 2003; see Appendix A). Fortunately, for most species in which 3D orientations and rotations are studied (birds, pinnipeds, sea turtles, in contrast to cetaceans), the device, even if somewhat tilted longitudinally and/or transversally, can easily be mounted in such a way that its surge axis lies in the body sagittal plane. One can then rely on a simplified procedure based on the relationship $\mathbf{B}^T = \mathbf{R}_2(\phi^*)\mathbf{R}_1(\lambda^*)\mathbf{D}^T$, with $\phi^* = \sin^{-1}(-a^*_U/\|\mathbf{a}^*\|)$ and $\lambda^* = \operatorname{atan}_2(-a^*_V, -a^*_W)$, where $\mathbf{a}^* = (a^*_U, a^*_V, a^*_W)$ is the acceleration vector (see §2) recorded by the device mounted on the back of the animal kept level ($\phi \approx \lambda \approx 0$) before release, i.e.

$$B_U = (D_U\,(a^{*2}_V + a^{*2}_W)^{0.5} - a^*_U\,(D_V\,a^*_V + D_W\,a^*_W)/(a^{*2}_V + a^{*2}_W)^{0.5})/\|\mathbf{a}^*\| \tag{1a}$$

$$B_V = (D_W\,a^*_V - D_V\,a^*_W)/(a^{*2}_V + a^{*2}_W)^{0.5} \tag{1b}$$

$$B_W = -(D_U\,a^*_U + D_V\,a^*_V + D_W\,a^*_W)/\|\mathbf{a}^*\| \tag{1c}$$

These matrices also apply to any unit vector that specifies an orientation in 3D space. This vector is usually initially expressed in terms of two angular spherical coordinates, $\xi \in [-\pi, \pi]$ and $\psi \in [-\pi/2, \pi/2]$. To be multiplied by a rotation matrix, it has first to be expressed in the equivalent Cartesian form, $(\cos(\xi)\cos(\psi), \sin(\xi)\cos(\psi), \sin(\psi))$, which, for simplicity, will be written hereafter with the compact notation $[\xi \bullet \psi]$ (i.e. $[\xi \bullet \psi]^T = \mathbf{R}_3(\xi)\mathbf{R}_2(\psi)\,(1, 0, 0)^T$).

## 2. Estimating posture and activity from 3D acceleration data

Accelerometers are sensitive to any type of linear acceleration, including gravity. An animal-borne high-frequency tri-axial accelerometer therefore makes it possible to determine both the body posture (i.e. orientation with respect to gravity) and the activity level, once the 'static' (gravity-induced) and 'dynamic' (movement-induced) components have been disentangled.

Call $\mathbf{a} = (a_U, a_V, a_W)$ the acceleration vector recorded at any given time $t$, where $a_U$, $a_V$, and $a_W$ corresponds to the values obtained for the surge, sway and heave axes of the accelerometer, respectively. When this device is at rest, one gets $\|\mathbf{a}\| = g$ (9.81 m/s$^2$), while the value obtained for each axis ranges between $-g$ and $g$, depending on its orientation ($-g$ when the axis considered is kept vertically upwards, $g$ when it is kept vertically downwards, 0 when it is kept horizontal). When the device is borne by a moving animal, the data obtained for each axis corresponds to the combination of a static and a dynamic components. Assuming that dynamic acceleration involves mainly fluctuations that tend to quickly counter-balance each other, the static acceleration at time $t$ can be accounted for by the mean vector $\bar{\mathbf{a}} = (\bar{a}_U, \bar{a}_V, \bar{a}_W)$, where $\bar{a}_U$, $\bar{a}_V$, and $\bar{a}_W$ are the mean values obtained for each axis separately, by averaging the values recorded between times $t-\Delta t/2$ and $t+\Delta t/2$. The temporal window $\Delta t$. must be long enough to smooth acceleration fluctuations but short enough to keep the best possible resolution. A width ranging between 1 and 3 s (depending on the species; Shepard et al. 2008a) is usually appropriate. The norm of the static acceleration, $\|\bar{\mathbf{a}}\| = (\bar{a}_U^2 + \bar{a}_V^2 + \bar{a}_W^2)^{0.5}$, should usually remain close to $g$ (otherwise the temporal window $\Delta t$ has to be enlarged). As it cannot be mounted at the body mass center, an accelerometer is also subjected to linear accelerations due to body rotations, which may result in underestimating the norm of the static acceleration $\|\bar{\mathbf{a}}\|$ (in contrast, the mean norm $\overline{\|\mathbf{a}\|}$ should remain close to $g$). However, the device is usually mounted sufficiently close to the body mass center to consider that values of $a_U$, $a_V$ and $a_W$ obtained reflect purely linear movements.

As the device may have been mounted tilted longitudinally by $\phi^*$ and/or transversally by $\lambda^*$, the body posture may differ from the device's posture. With the device mounted in such a way that its surge axis lies in the body's sagittal plane, the static acceleration vector of the freely moving animal, $\mathbf{A} = (A_U, A_V, A_W)$, can be derived from the static acceleration vector of the device, $\overline{\mathbf{a}}$, based on the simple relationship $\mathbf{A}^\mathsf{T} = \mathbf{R}_2(\phi^*)\mathbf{R}_1(\lambda^*)\overline{\mathbf{a}}^\mathsf{T}$ (i.e. Eqs (**1**) with $B$ replaced by $A$ and $D$ replaced by $\overline{a}$). Note that the norm is not affected by the tilt: $\|\mathbf{A}\| = \|\overline{\mathbf{a}}\|$. The body posture can then be specified by the vertical orientations, measured from the horizontal plane, of its surge axis, $\phi = \sin^{-1}(-A_U/\|\mathbf{A}\|)$, sway axis, $\phi' = \sin^{-1}(-A_V/\|\mathbf{A}\|)$, and heave axis, $\phi'' = \sin^{-1}(-A_W/\|\mathbf{A}\|)$. As these three axes are orthogonal to each other, the body posture can be specified unequivocally by only two angles, which correspond to the two angular spherical coordinates of a unit vector whose Cartesian coordinates are $\sin(\phi)$, $\sin(\phi')$ and $\sin(\phi'')$. The body posture can thus be expressed in terms of overall inclination,

$$\zeta = \pi/2 - \phi'' = \cos^{-1}(-A_W/\|\mathbf{A}\|) \tag{2a}$$

which measures the angular deviation between the heave axis and the $Z$ (upward vertical) axis, and postural orientation,

$$\chi = \operatorname{atan}_2(\sin(\phi'), \sin(\phi)) = \operatorname{atan}_2(-A_V, -A_U), \tag{2b}$$

which specifies in which body-related direction (forwards-backwards and/or rightwards-leftwards) the angular deviation between the heave and $Z$ axes occurs. This handy representation, based on $[\chi \bullet \phi''] = (\sin(\phi), \sin(\phi'), \sin(\phi''))$, is comparable to a spherical bull's eye spirit level whose equatorial plane stands for the animal's coronal plane and bubble stands for the $Z$ axis direction. It is however often more convenient, for subsequent computations, to specify the posture in terms of elevation angle,

$$\phi = \sin^{-1}(-A_U/\|\mathbf{A}\|) = \tan^{-1}(-A_U/(A_V^2 + A_W^2)^{0.5}), \tag{3a}$$

and bank angle, which corresponds to the orientation of the sway axis measured in the frontal plane from the horizontal level,

$$\lambda = \operatorname{atan}_2(\sin(\phi'), \sin(\phi'')) = \operatorname{atan}_2(-A_V, -A_W), \tag{3b}$$

based on $[\lambda \bullet \phi] = (\sin(\phi''), \sin(\phi'), \sin(\phi))$. The overall body inclination can be easily derived from the elevation and bank angles through the relationship $\cos(\zeta) = \cos(\phi) \cos(\lambda)$.

An accelerometer being an inertial measurement device, any movement-induced acceleration is recorded as the result of an apparent force acting in the reverse direction: backward for a forward acceleration, forward for a backward acceleration (i.e. deceleration), centrifugal for the centripetal acceleration involved in turning. The posture estimates obtained using Eqs (**2**) and (**3**) are expressed with respect to a reference plane that is orthogonal to this apparent force. As movement-induced accelerations tend to counter-balance each other over $\Delta t$, only gravity should matter ($\|\mathbf{A}\| \approx g$) and the reference plane remains (roughly) horizontal. However, this plane will be somewhat tilted when an animal undertakes a strong long-lasting forward, backward or centripetal acceleration (e.g., a bird taking off, landing, or gliding along a circle arc, respectively; e.g. Williams et al. 2015). Determining the body posture in this context is quite complex, and requires that the trajectory is known. Just to get a flavor, consider an animal that constantly accelerates for a time $> \Delta t$ in the horizontal plane (i.e., orthogonally to gravity), either tangentially or centripetally. In this simple example, one gets $\|\mathbf{A}\| \approx (g^2+h^2)^{0.5}$, with $h = \Delta v/\Delta t$ for a tangential acceleration, or $h = v^2/r$ for a centripetal acceleration, where $v$ is the linear speed and $r$ is the radius of curvature. The reference plane is tilted by $\tan^{-1}(h/g)$ with respect to the horizontal plane, either longitudinally for a tangential acceleration or transversally for a centripetal acceleration. The bank angle has then to be corrected: $\phi = \sin^{-1}(-A_U/\|\mathbf{A}\|) - \tan^{-1}(\Delta v/(g\Delta t))$ in the former case, $\lambda = \text{atan}_2(-A_V, -A_W) \pm \tan^{-1}(v^2/(gr))$ (– for a left turn, + for a right turn) in the latter case. Thus, for a turn at constant speed in the horizontal plane, a car with stiff suspensions will show no banking ($\lambda = 0$) while it is submitted to a strong transversal acceleration ($|A_V| = v^2/r$), whereas a bicycle has to adopt the bank angle $\lambda = \pm\tan^{-1}(v^2/(gr))$ to turn as expected while it is submitted to no noticeable transversal acceleration ($A_V \approx 0$). Analyses of the flights of a griffon vulture and an osprey (Olivier Duriez, unpublished data) showed that, unsurprisingly, the 'bicycle way' applies to soaring birds that loop in a thermal to gain altitude.

The data acquired with the accelerometer can also be used to assess activity level as the mean of the norm, $\|\mathbf{d}\| = (d_U^2+d_V^2+d_W^2)^{0.5}$, of the vector difference, $\mathbf{d} = (d_U, d_V, d_W) = \mathbf{a} - \bar{\mathbf{a}}$. The value of $\|\mathbf{d}\|$ will be referred hereafter to as the 'Dynamic Body Acceleration' (DBA). The DBA value computed in this way does not depend on the posture. It can therefore be considered a reliable measure of activity, including when the device is mounted on a collar (i.e. when the device's posture may be markedly different from the body posture). It was initially called (somewhat misleadingly, as it is a scalar) 'vectorial DBA' (veDBA) by Qasem et al. (2012), to distinguish it from the previous, mathematically inconsistent, way to combine the three $\mathbf{d}$ components, $|d_U|+|d_V|+|d_W|$, called 'overall DBA' (ODBA; R. Wilson et al. 2006), which results in unreliable (posture-dependent) activity values. On average, DBA is proportional to oxygen consumption and therefore can act as a suitable proxy for energetic expenses (Gleiss et al. 2011, Qasem et al. 2012; R. Wilson et al. 2020a). It can also provide a proxy for the linear speed (see §8).

Static and dynamic acceleration components, $\bar{\mathbf{a}}$ and $\mathbf{d}$, have initially to be computed at the same high frequency $f$ as the frequency with which raw data, $\mathbf{a}$, is acquired by using a *sliding* $\Delta t$-width window, with $\bar{\mathbf{a}}$ being attributed to the center of the window, which encompasses ca. $k = \text{round}(f\Delta t)$ values: $\bar{\mathbf{a}}$ is computed by averaging, for each axis separately, the central $a$ value, the $(k–1)/2$ (if $k$ is odd) or either $k/2$ or $(k–2)/2$ (if $k$ is pair), preceding and following $a$ values. Nevertheless, the amount of data to process in subsequent analyses can be dramatically reduced (divided ca. $k$) without significant loss of information by keeping, for each time step $\Delta t$, a single $\bar{\mathbf{a}}$ through subsampling (i.e. as if $\bar{\mathbf{a}}$ was computed using a *jumping* rather than *sliding* window), and a single DBA value through averaging (see §9).

## 3. Estimating heading from 3D accelero-magnetic data

Heading in 3D space is defined by both azimuth, θ, and elevation angle, φ. While the latter only requires a high-frequency tri-axial accelerometer to be estimated (Eq. **3**a), the former requires a high-frequency tri-axial accelero-magnetometer. Indeed, a tri-axial magnetometer, which records the geomagnetic field in 3D space, makes it possible to compute the azimuth with respect to the magnetic North only if its posture is known. For this purpose, it must be tightly associated with an accelerometer (on the same chip, sharing the same axes, and being synchronized with a common clock, but not necessarily working at the same frequency).

A magnetometer must be properly calibrated: a shift of the range of the values obtained (offset), specific to each axis, is expected because the device itself generates a magnetic field (hard iron effect). The sensitivity for each axis may also be differently affected by the local environment (soft iron effect). The raw values must therefore be corrected to obtain calibrated values expressed in the same 0-centered range for the three axes. The range width by itself does not matter. Call ($N_U$, $N_V$, $N_W$), ($E_U$, $E_V$, $E_W$), ($S_U$, $S_V$, $S_W$), and ($W_U$, $W_V$, $W_W$), the raw magnetic values measured for the surge ($U$), sway ($V$) and heave ($W$) axes when the surge axis, kept horizontal, points successively towards magnetic North, East, South and West, respectively. These values must be acquired twice, through two horizontal rotations of the surge axis: (i) with the device that rests on its bottom ($a_U \approx a_V \approx 0$, $a_W \approx -g$), and (ii) with the device that rests on its left or right flank ($a_U \approx a_W \approx 0$, $a_V \approx \pm g$). The 'bottom rotation' should let the $W$-indexed values unchanged and result in $E_U \approx W_U \approx (N_U+S_U)/2$ and $N_V \approx S_V \approx (E_V+W_V)/2$, whereas the 'flank rotation' should let the $V$-indexed values unchanged and result in $N_W \approx S_W \approx (E_W+W_W)/2$ and in $U$-indexed values alike those obtained with the 'bottom rotation' (otherwise, swap the labels of the axes). The raw magnetic values, $r_U$, $r_V$ and $r_W$ can then be translated into the calibrated ones:

$$m_U = (2r_U - N_U - S_U)/(N_U - S_U) \qquad \text{based on the 'bottom rotation'} \qquad (\textbf{4}\text{a})$$

$$m_V = (2r_V - E_V - W_V)/(E_V - W_V) \qquad \text{based on the 'bottom rotation'} \qquad (\textbf{4}\text{b})$$

$$m_W = \text{sgn}(a_v)(2r_W - E_W - W_W)/(E_W - W_W) \quad \text{based on the 'flank rotation'} \qquad (\textbf{4}\text{c})$$

Thus, for any axis rotated horizontally, the $m$ value varies from −1 (when the axis considered points southwards) to 1 (when it points northwards). Rescaling it by $B_m/\cos(\varphi_m)$, where $B_m$ is the local geomagnetic field intensity and $\varphi_m$ is the geomagnetic dip angle (inclination), to determine the actual geomagnetic value is unnecessary because the azimuth estimate rests on the ratio of two calibrated values (see below), which is independent of any rescaling factor. If the identities and directions of the three axes are known, a simpler calibration procedure consists in determining the minimum and maximum raw values, $r_{\min}$ and $r_{\max}$, for each axis separately. Such values can be determined in 3D by rotating the device at random for a while. However, it is not so easy to explore 3D space evenly. One can alternatively perform a full turn in the horizontal plane with the device set level ($a_U \approx a_V \approx 0$) to determine the horizontal $r_{\min}$ and $r_{\max}$ values for the surge and sway axes, and a further horizontal full turn with the device tilted on its right, left, front or rear flank ($a_W \approx 0$) to determine the horizontal $r_{\min}$ and $r_{max}$ values for the heave axis. For a given axis, any raw value $r$ can then be translated into the calibrated one $m = s(2r-r_{min}-r_{max})/(r_{max}-r_{min})$, with $s = 1$ or $s = -1$ depending on whether $r_{max}$ is obtained when the axis considered points northwards or southward, respectively.

Call $\bar{\mathbf{m}} = (\bar{m}_U, \bar{m}_V, \bar{m}_W)$ the mean calibrated magnetic vector obtained by averaging the magnetic values acquired, for each axis separately, over a short time interval $\Delta t$. With the device mounted with its surge axis lying in the body sagittal plane, the body-related mean magnetic vector $\mathbf{M} = (M_U, M_V, M_W)$ can be derived from the device-related one $\bar{\mathbf{m}}$ as $\mathbf{M}^T = \mathbf{R}_2(\phi^*)\mathbf{R}_1(\lambda^*)\bar{\mathbf{m}}^T$ (i.e. Eqs **1** with $B$ replaced by $M$ and $D$ replaced by $\bar{m}$). The body azimuth can then be assessed by combining the body-related static acceleration vector $\mathbf{A}$ and mean magnetic vector $\mathbf{M}$. Call $\underline{\mathbf{M}} = (\underline{M}_U, \underline{M}_V, \underline{M}_W)$ the mean magnetic vector that would have been obtained if the body was leveled while keeping its azimuth $\theta$ unchanged: $\underline{\mathbf{M}}^T = \mathbf{R}_2(\phi)\mathbf{R}_1(\lambda)\mathbf{M}^T$, where $\phi$ and $\lambda$ are the body elevation and bank angles (Eqs **3**). The body azimuth, measured positively counter-clockwise with respect to magnetic East, is $\theta = \text{atan}_2(\underline{M}_U, \underline{M}_V)$, i.e.

$$\theta = \text{atan}_2((A_V^2+A_W^2)M_U - A_U(A_VM_V+A_WM_W), (A_VM_W - A_WM_V)\|\mathbf{A}\|) \qquad (5)$$

An alternative formulation is $\theta = \theta^d - \text{atan}_2(c_2, c_1)$, where $\theta^d$ is the azimuth of the device (see Eq. **18**), and $c_1$ and $c_2$ are derived from the relationship $(c_1, c_2, c_3)^T = \mathbf{R}_2(\phi)\mathbf{R}_1(\lambda)[0 \bullet \phi^*]^T$. The 'true' azimuth value (i.e. based on geographical East) is obtained by subtracting the geomagnetic declination (which can however be ignored for fine-scale movement reconstruction; see §8). In the particular case of an animal that heads up or down ($\phi \approx \pm \pi/2$, $A_V \approx A_W \approx 0$), Eqs (**3**b) and (**5**) result in irrelevant values that vary randomly. The 3D body orientation can then be consistently specified by pseudo-azimuth $\theta = \theta' - \pi/2$, where $\theta' = \text{atan}_2(M_V, -\text{sgn}(A_U)M_W)$ is the direction of the sway axis, elevation angle $\phi = -\text{sgn}(A_U)\pi/2$, and pseudo-bank angle $\lambda = 0$. In some studies (e.g. R. Wilson et al. 2020b), one may need to express the orientation of the head relatively to the body. The required formulae are given in Appendix B.

By the way, note that the body-related mean magnetic vector $\mathbf{M}$ can also be used to compute magnetic 'elevation' and 'bank' angles as $\sin^{-1}(-M_U/\|\mathbf{M}\|)$ and $\text{atan}_2(-M_V, -M_W)$, which are insensitive to the dynamic accelerations that may bias the computation of the standard (gravity-based) elevation and bank angles, $\phi$ and $\lambda$. However, the geomagnetic field being characterized by a dip angle $\phi_m$, the reference plane (orthogonal to $\mathbf{M}$), against which the magnetic posture is specified, is tilted by $\pi/2-|\phi_m|$ with respect to the horizontal plane, making the magnetic posture hard to interpret. The standard posture can yet be assessed using $\mathbf{M}$ in the particular case of a soaring bird looping in a thermal (Williams et al. 2017, 2018). At each loop, the arrowhead of $\mathbf{M}$ draws a circle around the $Z$ axis with a radius of $\|\mathbf{M}\|\cos(\phi_m)$. The sum $\mathbf{\Sigma M}$ of the $\mathbf{M}$ vectors sampled evenly over a loop is therefore vertical and deviates from the heave axis by $\zeta$ ($\phi_m < 0$) or $\pi - \zeta$ ($\phi_m > 0$), with $\zeta = \cos^{-1}(|\Sigma M_W|/\|\mathbf{\Sigma M}\|)$. The elevation angle, $\phi = \sin^{-1}(-\text{sgn}(\phi_m)\Sigma M_U/\|\mathbf{\Sigma M}\|)$, should be null (a soaring bird is expected to gain altitude only thanks to the vertical advection of warm air) whereas the bank angle, $\lambda = \tan^{-1}(-\text{sgn}(\phi_m) \Sigma M_V/|\Sigma M_W|)$, should be equal to $\pm \tan^{-1}(v^2/(gr))$, where $v$ is the linear speed and $r$ is the radius of curvature (a soaring bird is expected to turn in bending over as a bicycle does).

## 4. Estimating body rotations from 3D gyroscopic data

A high-frequency tri-axial gyrometer measures the angular velocity components in the coronal, sagittal and frontal planes. They are referred to as yaw, pitch and roll speeds, $\dot{\alpha}$, $\dot{\beta}$ and $\dot{\rho}$, respectively. To be consistent with the ENU convention, $\dot{\alpha}$ and $\dot{\rho}$ are positive but $\dot{\beta}$ is negative for a rotation that appears anticlockwise when the corresponding rotation axis points towards the observer (the convention used in the firmware of the device may be different). With the device mounted in such a way that its surge axis lies in the body's sagittal plane, the body angular velocity $(\dot{\rho}, \dot{\beta}, \dot{\alpha})$ can be derived from the recorded one $(\dot{\rho}^d, \dot{\beta}^d, \dot{\alpha}^d)$ based on the relationship $(\dot{\rho}, -\dot{\beta}, \dot{\alpha})^T = \mathbf{R}_2(\phi^*)\mathbf{R}_1(\lambda^*)(\dot{\rho}^d, -\dot{\beta}^d, \dot{\alpha}^d)^T$ (i.e. Eqs (**1**) with $B_U = \dot{\rho}$, $B_V = -\dot{\beta}$, $B_W = \dot{\alpha}$, $D_U = \dot{\rho}^d$, $D_V = -\dot{\beta}^d$, and $D_W = \dot{\alpha}^d$). The body rotations yaw, pitch and roll for a given time step $\Delta t = t_i - t_{i-1}$ can be computed simply as $\alpha_i = \Sigma_j \dot{\alpha}_{i,j} / f$, $\beta_i = \Sigma_j \dot{\beta}_{i,j} / f$ and $\rho_i = \Sigma_j \dot{\rho}_{i,j} / f$, for $j = 1 \ldots k$, based on the $k = \text{round}(f\Delta t)$ angular speeds, $\dot{\alpha}_{i,j}$, $\dot{\beta}_{i,j}$ and $\dot{\rho}_{i,j}$ measured with the frequency $f$.

A high-frequency tri-axial gyrometer is the most reliable device for measuring yaw, pitch and roll. Without it, these three body rotations can be assessed indirectly from data acquired by a high-frequency tri-axial accelero-magnetometer (see §6). Reciprocally, changes of azimuth $\Delta\theta$ for a given time step $\Delta t$ can be estimated using a high-frequency tri-axial accelero-gyrometer (see Eq. **13**). However, estimating the current azimuth as the sum of changes of azimuth and initial azimuth $(\theta_i = \Sigma_{j=1}^i \Delta\theta_j + \theta_0)$ quickly accumulates random errors. Hence, accelero-gyrometers may be used in place of accelero-magnetometers to assess azimuth only at very short term. A high-frequency accelero-magneto-gyrometer can be useful to cross-check heading estimates, but involves a non-negligible battery drain. It may help infer high-speed movements with great accuracy (A. Wilson et al. 2013), in particular because the tilting of the reference plane when the animal undertook strong long-lasting accelerations makes difficult the estimation of the posture based only on acceleration data (see §2).

## 5. Changes of heading and posture

A given heading can be represented as a point at the surface of a unit sphere centered on the animal's current location and whose reference axes $(X, Y, Z)$ match the geographical ones (East, North, Up). The change of heading is done in the plane defined by the two unit vectors, $(\theta_{i-1}, \phi_{i-1})$ and $(\theta_i, \phi_i)$, standing for heading at two successive time steps. It can be represented by the orthodromic arc that links the two corresponding points on the unit sphere (Fig. 2). This is the minor arc of the 'great circle' (i.e. the largest circle that can be drawn on a sphere) passing through the two points considered, and corresponds to the shortest possible surface route (geodesic) between them. The change of heading is therefore characterized by the arc size, $\omega_i \in [0, \pi]$, and initial inclination, $\delta_i^i \in [-\pi, \pi]$, measured positively counterclockwise from the horizontal level in the animal's frontal plane (i.e. tangentially to the unit sphere) at heading $(\theta_i, \phi_{i-1})$. Assuming that, for the short time interval $\Delta t = t_i - t_{i-1}$, the change of heading from $(\theta_{i-1}, \phi_{i-1}) = (\tilde{\theta}_i(0), \tilde{\phi}_i(0))$ to $(\theta_i, \phi_i) = (\tilde{\theta}_i(1), \tilde{\phi}_i(1))$ was done at constant speed $\omega_i/\Delta t$, the heading ($\tilde{\theta}_i(q)$, $\tilde{\phi}_i(q)$) at time $t_{i-1}+q\Delta t$ with $q \in [0, 1]$ is obtained from the relationship $[\tilde{\theta}_i(q) \bullet \tilde{\phi}_i(q)]^T = \mathbf{R}_3(\theta_{i-1})\mathbf{R}_2(\phi_{i-1})\mathbf{R}_1(\delta_i^i)[q\omega_i \bullet 0]^T$:

$$\tilde{\theta}_i(q) = \theta_{i-1} + \text{atan}_2(\cos(\delta_i^i), \cos(\phi_{i-1})/\tan(q\omega_i) - \sin(\delta_i^i)\sin(\phi_{i-1})) \tag{6a}$$

$$\tilde{\phi}_i(q) = \sin^{-1}(\sin(q\omega_i)\sin(\delta_i^i)\cos(\phi_{i-1}) + \cos(q\omega_i)\sin(\phi_{i-1})) \tag{6b}$$

Reciprocally, the size, $\omega_i$, and the local inclination, $\tilde{\delta}_i(q)$ with $q \in [0, 1]$, of the orthodromic arc representing the change of heading between times $t_{i-1}$ and $t_i$ can be computed as:

$$\omega_i = \cos^{-1}(\sin(\phi_{i-1})\sin(\phi_i) + \cos(\phi_{i-1})\cos(\phi_i)\cos(\theta_i - \theta_{i-1})) \tag{7a}$$

$$\tilde{\delta}_i(q) = \text{atan}_2(\cos(q\omega_i)\sin(\phi_i) - \cos((1-q)\omega_i)\sin(\phi_{i-1}),$$
$$\cos(\phi_{i-1})\cos(\phi_i)\sin(\theta_i - \theta_{i-1})) \tag{7b}$$

Eqs (**6**) and (**7**) can also be partly derived from the spherical law of cosines. Note that, while the elevation angle $\tilde{\phi}_i(q)$ varies from $\phi_{i-1} = \tilde{\phi}_i(0)$ to $\phi_i = \tilde{\phi}_i(1)$, and the local reorientation inclination $\tilde{\delta}_i(q)$ varies from initial inclination, $\delta_i^i = \tilde{\delta}_i(0)$, to final inclination, $\delta_i^f = \tilde{\delta}_i(1)$, the product $p_i = \cos(\tilde{\phi}_i(q))\cos(\tilde{\delta}_i(q))$, which characterizes the inclination $\hat{\phi}_i = \cos^{-1}(|p_i|)$ of the reorientation plane for the time step $\Delta t = t_i - t_{i-1}$, remains constant ($\hat{\phi}_i$ corresponds therefore to the elevation angle at the apex of the great circle to which the orthodromic arc belongs, reached at azimuth $\hat{\theta}_i = \tilde{\theta}_i(q) + \text{sgn}(\tan(\tilde{\delta}_i(q)))\cos^{-1}(\tan(\tilde{\phi}_i(q))/\tan(\hat{\phi}_i)))$.

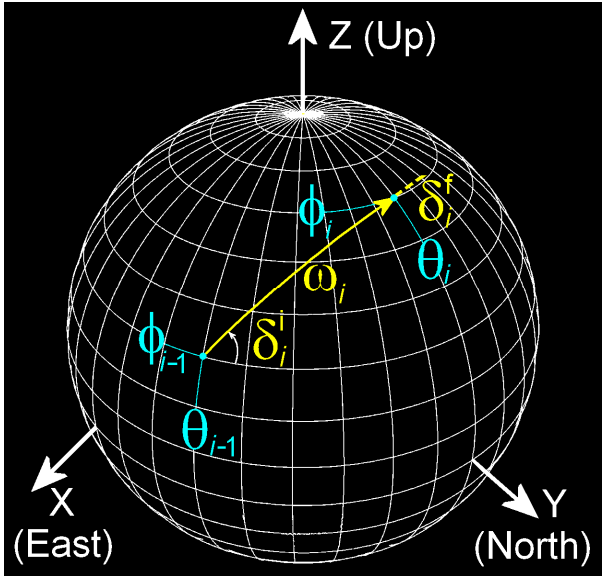

**Fig. 2**. **Change of heading from ($\theta_{i-1}$, $\phi_{i-1}$) to ($\theta_i$, $\phi_i$).** Modeled as the orthodromic arc linking these two headings, this reorientation can be suitably represented on a unit sphere in a mixed body-bound/Earth-bound frame of reference: the sphere is centered on the animal's location $\mathbf{x}_{i-1} = (x_{i-1}, y_{i-1}, z_{i-1})$, but its reference directions are provided by the $X$ axis pointing eastwards, the $Y$ axis pointing northwards and the $Z$ axis pointing upwards (ENU convention). Azimuths, $\theta_{i-1}$ and $\theta_i$, are equivalent to longitudes while elevation angles, $\phi_{i-1}$ and $\phi_i$, are equivalent to latitudes. The magnitude of the change of heading, i.e. the arc size $\omega_i$, is an 'internal' angle whose vertex corresponds to the center of the unit sphere, while the initial and final arc inclinations, $\delta_i^i$ and $\delta_i^f$, are 'surface' angles measured in the animal's current frontal plane, i.e. the plane orthogonal to the sphere at their respective vertices. In this example of a counter-clockwise reorientation, the values of the different angles are $\theta_{i-1} = \phi_{i-1} = \pi/6$, $\theta_i = \pi/2$, $\phi_i = \pi/3$, $\omega_i \approx \pi/3.637$, $\delta_i^i \approx \pi/3.256$ and $\delta_i^f \approx \pi/19.023$ (replace $\pi$ by 180 to obtain values in degrees). The product $p_i = \cos(\phi_{i-1})\cos(\delta_i^i) = \cos(\phi_i)\cos(\delta_i^f)$ characterizes the inclination (the inclination of the reorientation plane, $\hat{\phi}_i = \cos^{-1}(|p_i|) \approx \pi/2.978$. It corresponds to the elevation angle at the highest point of the great circle to which the orthodromic arc belongs, reached at azimuth $\pi/1.784$. For a counter-clockwise reorientation, $|\delta_i^i|$ and $|\delta_i^f|$ cannot be larger than $\hat{\phi}_i$ while, for a clockwise reorientation, they cannot be smaller than $\pi - \hat{\phi}_i$, with $\hat{\phi}_i \in [0, \pi/2]$.

A key point is that the change of local arc inclination from $\delta_i^i$ to $\delta_i^f$ compels the bank angle to change passively (i.e. without involving any roll) by $\delta_i^f - \delta_i^i$. If some roll occurs concomitantly, it simply adds to this passive change. Accordingly, the local arc inclination expressed with respect to the coronal plane, which ranges from $\eta_i^i = \delta_i^i - \lambda_i$ at time $t_{i-1}$ to $\eta_i^f = \delta_i^f - \lambda_i$ at time $t_i$, remains constant ($\eta_i^f = \eta_i^i$) during any change of heading without rolling.

Assuming that roll, $\rho_i = \eta_i^i - \eta_i^f \in [-\pi, \pi]$, is done at constant speed $\rho_i/\Delta t$ between times $t_{i-1}$ and $t_i$, the bank angle at any intermediate time $t_{i-1}+q\Delta t$ with $q \in [0, 1]$ can be interpolated as

$$\tilde{\lambda}_i(q) = \lambda_{i-1} + \tilde{\delta}_i(q) - \delta_i^i + q(\eta_i^i - \eta_i^f) \tag{8}$$

Finally, it is worth noting that Eq. (7a) reflects that the dot product of two unit vectors is simply equal to the cosine of angular discrepancy between them: $\cos(\omega_i) = [\theta_{i-1} \bullet \phi_{i-1}] \bullet [\theta_i \bullet \phi_i]$. Similarly, the cosine of the size, $\varpi_i$, of the change of posture can be computed as

$$\cos(\varpi_i) = [\lambda_{i-1} \bullet \phi_{i-1}] . [\lambda_i \bullet \phi_i] = [\chi_{i-1} \bullet \phi_{i-1}''] \bullet [\chi_i \bullet \phi_i''] = \mathbf{A}_{i-1}\mathbf{A}_i/(\|\mathbf{A}_{i-1}\|\|\mathbf{A}_i\|) \tag{9}$$

## 6. Relating body rotations to changes of heading and posture

In 2D space, posture is irrelevant: the heading is defined only by the azimuth, and the single possible type of rotation, a change of azimuth, is usually referred to as the turning angle. It is done in the animal's coronal plane and therefore corresponds to yaw. In 3D space, heading is defined by both azimuth and elevation angle. The change of heading is done in a plane that may be quite different from the animal's coronal plane, and therefore may involve both yaw and pitch. Furthermore, even if some roll does not change the heading, it involves a change in the orientation of the coronal plane with respect to the horizontal plane, and therefore affects how the change of heading can be decomposed in terms of yaw and pitch (Fig. 3).

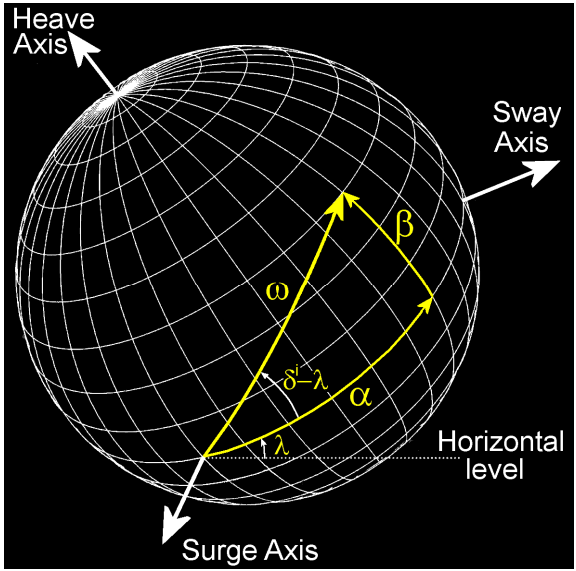

**Fig. 3. Decomposition of the change of heading in terms of yaw and pitch.** This reorientation can be suitably represented as an orthodromic arc drawn on a unit sphere in a pure body-bound frame of reference: the sphere is centered on the animal location, and the three reference axes correspond to the animal's main body axes. The rotation around the heave axis, i.e. yaw ($\alpha$), and the rotation around the sway axis, i.e. pitch ($\beta$), done sequentially in this order, correspond to the two orthogonal components of the orthodromic arc representing the change of heading. They are related to the size ($\omega$) and initial inclination ($\delta^i$) of this arc, and to the bank angle ($\lambda$), by the expression $[\alpha_i \bullet \beta_i]^T = \mathbf{R}_1(\delta^i - \lambda)[\omega_i \bullet 0]^T$ (Eqs **10**). In this sequential approach, roll is assumed to occur only before and after pitch (its effect is already included in the $\lambda$ value) so that the arc inclination expressed with respect to the animal's coronal plane remains constant (equal to $\delta^i - \lambda$) during the change of heading. Other formulae apply when yaw, pitch and roll are done simultaneously (see text and Fig.4). In this example, the values of the different angles $\alpha = \pi/3$, $\beta = \pi/6$, $\omega \approx \pi/2.80$, $\delta^i - \lambda \approx \pi/5.34$ (replace $\pi$ by 180 to obtain values in degrees). Note that $\delta^i$ and $\lambda$ are 'surface' angles measured in the animal's frontal plane, i.e. the plane orthogonal to the surge axis, while $\alpha$, $\beta$ and $\omega$ are 'internal' angles whose vertex corresponds to the center of the unit sphere.

Yaw corresponds to a change of azimuth ($\alpha_i = \omega_i = \theta_i - \theta_{i-1}$) for a turn by $\alpha_i$ around the heave axis fixed vertically ($\phi_{i-1} = \phi_i = \lambda_{i-1} = \lambda_i = \beta_i = \rho_i = 0$). Pitch corresponds to a change of elevation angle ($\beta_i = \omega_i = \phi_i - \phi_{i-1}$ for a turn by $\beta_i$ around the sway axis fixed horizontally without going beyond the vertical ($\lambda_{i-1} = \lambda_i = 0$, $\theta_i = \theta_{i-1}$, $\alpha_i = \rho_i = 0$). Roll corresponds to a change of bank angle ($\rho_i = \lambda_i - \lambda_{i-1}$) for a turn by $\rho_i$ around the surge axis set to a fixed heading ($\theta_i = \theta_{i-1}$, $\phi_i = \phi_{i-1}$, $\alpha_i = \beta_i = \omega_i = 0$). In the general case, the posture at time $t_{i-1}$ can be any. Furthermore, a change of heading from time $t_{i-1}$ to time $t_i$ can generate a passive change (i.e. without roll; see Eq. **8** with $\eta_i^i = \eta_i^f$) of the bank angle. Consequently, yaw, pitch and roll are usually related to changes of heading/posture in a more complex way.

These three body rotations are expected to be done simultaneously. If they were done sequentially, with pitch done just after yaw and roll done only before yaw and after pitch, yaw and pitch could be considered the two orthogonal components of heading at time $t_i$ expressed in the frame of reference defined by the heading/posture at time $t_{i-1}$: $[\alpha_i \bullet \beta_i]^T = \mathbf{R}_1(-\lambda_{i-1})\mathbf{R}_2(-\phi_{i-1})\mathbf{R}_3(-\theta_{i-1})[\theta_i \bullet \phi_i]^T$, as well as the two orthogonal components of the orthodromic arc that represents the change of heading between times $t_{i-1}$ and $t_i$: $[\alpha_i \bullet \beta_i]^T = \mathbf{R}_1(\eta_i)[\omega_i \bullet 0]^T$, i.e.

$$\alpha_i = \text{atan}_2(\cos(\eta_i), \cot(\omega_i)) \tag{10a}$$

$$\beta_i = \sin^{-1}(\sin(\eta_i)\sin(\omega_i)), \tag{10b}$$

where $\eta_i = \delta_i^i - \lambda_{i-1} = \delta_i^f - \lambda_i$ is the value, kept constant (no concomitant roll), of the reorientation inclination with respect to the animal's coronal plane (Fig. 3). It determines how the change of heading is shared between yaw and pitch, from pure yaw ($\eta_i = 0$ or $|\eta_i| = \pi$) to pure pitch ($|\eta_i| = \pi/2$). This sequential decomposition of the change of heading proves useful to model 3D random walks in terms of yaw and pitch (Benhamou 2018). It suitably approximates what happens with simultaneous body rotations if the change of heading is sufficiently small to get $\sin(\omega_i) \approx \omega_i$, but generates biased results when $\omega_i$ is large, as may occur when heading is estimated every $\Delta t$ rather than at the high recording frequency $f$. For example, Eqs (**10**) cannot account for the perfect balance between yaw and pitch ($|\alpha_i| = |\beta_i|$) that is expected when these rotations are done simultaneously while the reorientation plane is at mid-way between the animal's coronal and sagittal planes (i.e. $|\eta_i| = \pi/4$ or $|\eta_i| = 3\pi/4$).

The simultaneous occurrence of yaw, $\alpha_i$, pitch, $\beta_i$, and roll, $\rho_i$, can be mimicked by virtually subdividing the orthodromic arc representing the change of heading/posture from $(\theta_{i-1}, \phi_{i-1}, \lambda_{i-1})$ to $(\theta_i, \phi_i, \lambda_i)$ in $k$ sub-arcs. This change can then be modeled as a series of $k$ blocks of three successive small rotations, done in this order: (1) a rotation around the surge axis by $\rho_{i,j}$, (2) a rotation around the heave axis by $\alpha_{i,j}$ and (3) a rotation around the sway axis by $\beta_{i,j}$, for $j = 1 \dots k$. The concomitant change of location from $\mathbf{x}_{i-1}$ to $\mathbf{x}_i$ can be accounted for by coupling the rotation by $\rho_{i,j}$ around the surge axis with a translation along this axis by $2r\sin(0.5\omega_i/k)$, which corresponds to the length of the chord of an arc of circle with radius $r = 0.5\|\mathbf{x}_i - \mathbf{x}_{i-1}\|/\sin(\omega_i/2)$ and length $r\omega_i/k$. The $k$ sub-arcs have the same size, $\omega_i/k$, and involve the same amount of roll, $\rho_{i,j} = \rho_i/k$ for any $j$, with $\rho_i = \eta_i^i - \eta_i^f$. Without roll (i.e. $\eta_i^i = \eta_i^f$), they are all characterized by the same yaw and pitch decomposition: $\alpha_{i,j} = \alpha_i/k$ and $\beta_{i,j} = \beta_i/k$ for any $j$. Otherwise, the incremental change by $-\rho_i/k$ of the reorientation inclination with respect to the animal's coronal plane, from $\eta_i^i$ to $\eta_i^f$, leads each sub-arc to be decomposed in its own way, which can be derived from Eqs (**10**): $\alpha_{i,j} = \tan^{-1}(\cos(\tilde{\eta}_i(q))\tan(\omega_i/k))$ and $\beta_{i,j} = \sin^{-1}(\sin(\tilde{\eta}_i(q))\sin(\omega_i/k))$, with $\tilde{\eta}_i(q) = \tilde{\delta}_i(q) - \tilde{\lambda}_i(q) = \eta_i^i - q\rho_i$ and $q = j/k$. When $k$ is sufficiently large, these expressions can be rewritten $\alpha_{i,j} \approx \cos(\tilde{\eta}_i(q))\ \omega_i/k$ and $\beta_{i,j} \approx \sin(\tilde{\eta}_i(q))\ \omega_i/k$ (small-angle approximation). Yaw and pitch for the whole arc can therefore be expressed as $\alpha_i = \Sigma_j\alpha_{i,j} = \overline{\cos(\tilde{\eta}_i(q))}\ \omega_i$ and $\beta_i = \Sigma_j\beta_{i,j} = \overline{\sin(\tilde{\eta}_i(q))}\ \omega_i$. If roll is null, i.e. $\tilde{\eta}_i(q) = \eta_i$ for any $q$, one gets $\alpha_i$ $= \cos(\eta_i)\,\omega_i$ and $\beta_i = \sin(\eta_i)\,\omega_i$, and therefore $|\alpha_i| = |\beta_i|$ for $|\eta_i| = \pi/4$ or $|\eta_i| = 3\pi/4$ as expected. Otherwise, with roll speed, $\dot\rho_i = \rho_i/\Delta t$ with $\rho_i = \eta_i^i - \eta_i^f \in [-\pi, \pi]$, considered constant for the short time step $\Delta t$, $\tilde\eta_i(q)$ is uniformly distributed between $\tilde\eta_i(0) = \eta_i^i = \overline{\eta}_i + \rho_i/2$ and $\tilde\eta_i(1) = \eta_i^f = \overline{\eta}_i - \rho_i/2$, where $\overline{\eta}_i = (\eta_i^i + \eta_i^f)/2 = \tilde\eta_i(0.5)$ is the mean value. Yaw and pitch for the time step $\Delta t = t_i - t_{i-1}$ can therefore be inferred as:

$$\alpha_i = (\sin(\eta_i^i) - \sin(\eta_i^f))\,\omega_i/\rho_i = 2\cos(\overline{\eta}_i)\,\sin(\rho_i/2)\,\omega_i/\rho_i \tag{11a}$$

$$\beta_i = (\cos(\eta_i^f) - \cos(\eta_i^i))\,\omega_i/\rho_i = 2\sin(\overline{\eta}_i)\,\sin(\rho_i/2)\,\omega_i/\rho_i \tag{11b}$$

based on the values of $\omega_i$, $\eta_i^i$ and $\eta_i^f$ derived from the headings/postures ($\theta_{i-1}$, $\phi_{i-1}$, $\lambda_{i-1}$) and ($\theta_i$, $\phi_i$, $\lambda_i$), estimated from the accelero-magnetic measures made between times $t_{i-1} - \Delta t/2$ and $t_{i-1} + \Delta t/2$ and between times $t_i - \Delta t/2$ and $t_i + \Delta t/2$, respectively (Fig. 4).

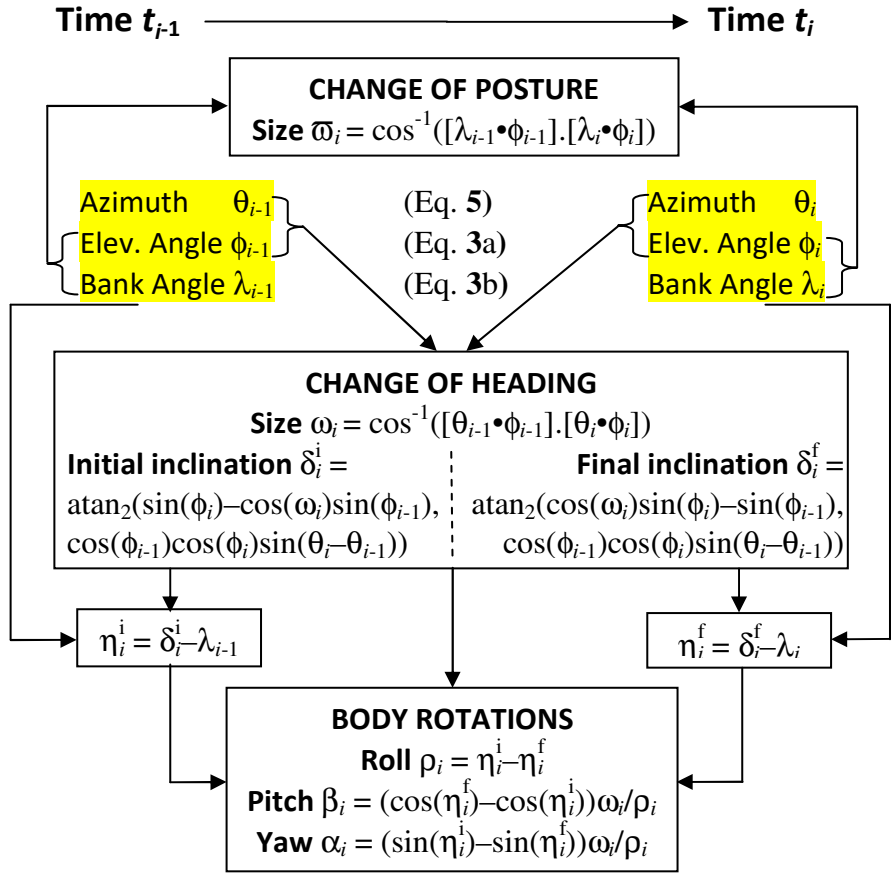

**Fig. 4. Inferring body rotations from time series of heading/posture.** An animal changes its heading (azimuth and elevation angle) and its posture (elevation and bank angles) through rotations – roll, pitch and yaw – done around its main three body axes – surge, sway and heave, respectively. These rotations can be inferred from the time series of successive headings and postures (expressed in the Earth-bound frame of reference specified by a geographical direction and gravity). The initial and final inclinations of the change of heading expressed with respect to the body coronal plane ($\eta_i^i$ and $\eta_i^f$) provides the interface between the body-bound and Earth-bound frames of reference that is required to back convert changes of orientations into rotations (with $\eta_i^i \approx \eta_i^f$, one gets $\alpha_i \approx \cos((\eta_i^i + \eta_i^f)/2)\omega_i$ and $\beta_i \approx \sin((\eta_i^i + \eta_i^f)/2)\omega_i)\rho_i$). Recall: $[\xi\bullet\psi]$ stands for $(\cos(\xi)\cos(\psi),\,\sin(\xi)\cos(\psi),\,\sin(\psi))$.

Reciprocally, the 3D reorientation for the time step $\Delta t = t_i - t_{i-1}$ can be assessed from the actual values of yaw, pitch, and roll derived from angular speeds measured by a high-frequency tri-axial gyrometer (see §4). With a time step $\Delta t$ sufficiently short to warrant that the angles obtained in this way remain in the range $[-\pi, \pi]$, the size of the change of heading and its mean inclination with respect to the animal's coronal plane can be computed as:

$$\omega_i = 0.5\rho_i(\alpha_i^2+\beta_i^2)^{0.5}/\sin(\rho_i/2) \tag{12a}$$

$$\overline{\eta}_i = \mathrm{atan}_2(\beta_i, \alpha_i) \tag{12b}$$

By coupling postures, $(\phi_{i-1}, \lambda_{i-1})$ and $(\phi_i, \lambda_i)$, estimated from the acceleration data recorded between times $t_{i-1}-\Delta t/2$ and $t_{i-1}+\Delta t/2$, and between times $t_i-\Delta t/2$ and $t_i+\Delta t/2$, respectively, and body rotations $\alpha_i$, $\beta_i$, and $\rho_i$ estimated from gyrometric data recorded between times $t_{i-1}$ and $t_i$, the change of azimuth between times $t_{i-1}$ and $t_i$, $\Delta\theta_i = \theta_i - \theta_{i-1}$, can therefore be computed as

$$\Delta\theta_i = \mathrm{atan}_2(\sin(\omega_i)(\cos(\phi_{i-1})\cos(\delta_i^i)+\cos(\phi_i)\cos(\delta_i^f))/2,\ \cos(\omega_i)-\sin(\phi_{i-1})\sin(\phi_i)) \tag{13}$$

with $\delta_i^i = \lambda_{i-1}+ \overline{\eta}_i +\rho_i/2$ and $\delta_i^f = \lambda_i+\overline{\eta}_i -\rho_i/2$, and $\omega_i$ and $\overline{\eta}_i$ derived from Eqs (**12**).

## 7. KDE-based computations of spherical distributions

A useful way to assess a distribution rests on kernel density estimation (KDE; Silverman 1986). KDE generates smoothed histograms that do not depend on the arbitrary choice of the grid origin and bin size, which may bias the representation obtained with standard histograms. Furthermore, building a histogram on a sphere involves a tricky tessellation (R. Wilson et al. 2016). To apply KDE to locations on the plane, a virtual horizontal grid with a small cell size is overlaid over the plane and the density is estimated at the center of each cell of the grid by setting an isotropic bivariate kernel, usually Gaussian, on each of the $n$ locations. The same principle applies to the locations on a unit sphere such as the ending points of 3D unit vectors $\boldsymbol{\mu}_i = (\xi_i, \psi_i)$ with $\xi_i \in [-\pi, \pi]$ (longitude-like) and $\psi_i \in [-\pi/2, \pi/2]$ (latitude-like). For this purpose, the horizontal planar grid is replaced by a spherical grid defined in terms of longitude-like and latitude-like value, and the isotropic bivariate Gaussian kernels are replaced by bivariate von Mises-Fisher kernels. In this way, KDE can apply to heading, $(\theta, \phi)$, change of heading, $(\delta^i, \omega)$, and posture, either $(\phi, \lambda)$ or $(\zeta, \chi)$, as azimuth, $\theta$, initial arc inclination, $\delta^i$, bank angle, $\lambda$, and postural orientation, $\chi$, behave as longitudes, elevation angle, $\phi$, behaves as a latitude, and arc size, $\omega$, and overall inclination, $\zeta$, behave as co-latitudes (i.e. $\pi/2 - \omega$ and $\pi/2 - \zeta$ behave as latitudes). Note however that this makes sense to compute a distribution (whichever the method used) only when the data is generated by a stationary process. Otherwise, the distribution obtained will arbitrarily depend on when the data considered begins and ends to be collected. It is therefore advised to check whether the mean and the variance obtained for the first, second, and third thirds of the time series of the data considered remain similar (Benhamou 2014) before attempting to build the distribution.

For any location $\mathbf{x}$ on the unit sphere, the $\boldsymbol{\mu}$-centered bivariate von Mises-Fisher probability density function with concentration parameter $\kappa$ is

$$f(\mathbf{x}) = \kappa \exp(\kappa \cos(\omega_{\mathbf{x},\boldsymbol{\mu}}))/(4\pi \sinh(\kappa)) \tag{14a}$$

where $\omega_{\mathbf{x},\boldsymbol{\mu}}$ is the size of the orthodromic arc $\widehat{\mathbf{x}\boldsymbol{\mu}}$, i.e. $\cos(\omega_{\mathbf{x},\boldsymbol{\mu}}) = \mathbf{x}.\boldsymbol{\mu} = 1 - (\mathbf{x}-\boldsymbol{\mu})^2/2$. The mean vector length of this distribution is $E(\cos(\omega_{\mathbf{x},\boldsymbol{\mu}})) = \coth(\kappa) - 1/\kappa$ (Fisher 1953). With $\kappa > 6$, one gets $\sinh(\kappa) \approx \cosh(\kappa) \approx \exp(\kappa)/2$, so that the von Mises-Fisher distribution can then be suitably approximated as:

$$f(\mathbf{x}) \approx \exp(-0.5(\mathbf{x}-\boldsymbol{\mu})^2/\sigma^2)/(2\pi\sigma^2) \qquad \textbf{(14}b)$$

with $\sigma^2 = 1/\kappa \approx 1-E(\cos(\omega_{\mathbf{x},\boldsymbol{\mu}}))$. Thus, a $\boldsymbol{\mu}$-centered bivariate von Mises-Fisher distribution with concentration parameter $\kappa > 6$ and a zero-centered isotropic bivariate Gaussian distribution with variance $\sigma^2 = 1/\kappa$ reflect each other through a $\boldsymbol{\mu}$-centered Lambert azimuthal equal-area projection, which maps orthodromic arcs originating at $\boldsymbol{\mu}$ as their corresponding chords. Consequently, the area encompassed within any given cumulative isopleth is the same for both distributions. For example, the area within the 0.95 cumulative isopleth for the former distribution with $\kappa=6$ or the latter distribution with $\sigma^2=1/6$ is equal to $\pi$ (corresponding to 1/4 of a unit sphere area or to a unit circle area, respectively).

The standard-deviation of a distribution used as a kernel in KDE is usually referred to as the smoothing parameter (or bandwidth) $h$. The value required to reliably estimate a spherical distribution is unlikely to be $>1/\sqrt{6}\approx\pi/8$. Consequently, Eq. (**14**b) should act as an appropriate kernel for building most (if not all) KDE-based spherical distributions. The density at the center $\mathbf{c} = (\xi_c, \psi_c)$ of a given cell of the spherical grid can therefore be estimated as

$$U_{\mathbf{c}} = \frac{1}{2\pi n h^2} \sum_{i=1}^{n} \exp\left(-\frac{(\mathbf{c}-\boldsymbol{\mu}_i)^2}{2h^2}\right) \qquad \textbf{(15)}$$

This expression is identical to the Gaussian kernel-based expression that applies to planar distributions if one agrees that, as on the plane, $\|\mathbf{c}-\boldsymbol{\mu}_i\|$ on the sphere is the Euclidean distance between $\mathbf{c}$ and $\boldsymbol{\mu}_i$ (i.e. the chord length of the arc $\overset{\frown}{\mathbf{c}\boldsymbol{\mu}_i}$ rather than the angular distance $\omega_{\mathbf{c},\boldsymbol{\mu}}$):

$$(\mathbf{c}-\boldsymbol{\mu}_i)^2 = 2(1 - \sin(\psi_c)\sin(\psi_i) - \cos(\psi_c)\cos(\psi_i)\cos(\xi_i-\xi_c)) \qquad \textbf{(16)}$$

When KDE applies to locations on a plane, $\mathbf{x}_i = (x_i, y_i)$ with $i = 1 \dots n$, that are independent of each other and come from an isotropic bivariate Gaussian distribution with standard deviation $\sigma$, the best trade-off between a level of smoothing sufficiently high to eliminate noise and sufficiently low to keep relevant information is given by $h_{ref} = \sigma/n^{1/6}$ (Silverman 1986). When dealing with 3D orientations $\boldsymbol{\mu}_i = (\xi_i, \psi_i)$, the corresponding value is $h_{ref} = \kappa^{-0.5}/n^{1/6} = (1-\|\Sigma_i[\xi_i\bullet\psi_i]\|/n)^{0.5}/n^{1/6}$ (see example in Fig. 5). However, as this double assumption never holds true, the $h_{ref}$ value only provides an order of magnitude of the value of smoothing parameter required for reliably representing the distribution.

The fraction of distribution attributed to a given $\mathbf{c}$-centered cell of the virtual spherical grid is equal to $a_{\mathbf{c}}U_{\mathbf{c}}$, where $a_{\mathbf{c}}$ is the cell area. If the grid is based on a regular binning of longitudes-like $\xi$ and latitudes-like $\psi$ with angular bin size $\upsilon < h$, the cell area is variable: $a_{\mathbf{c}} = 2\upsilon\sin(\upsilon/2)\cos(\psi_c) \approx \upsilon^2\cos(\psi_c)$. A grid with a constant cell area $a_{\mathbf{c}} = \upsilon s$ can be designed through a regular binning of longitudes-like and sines of latitudes-like, with bin sizes $\upsilon$ and $s$, respectively. To compute the smallest area encompassed within a given cumulative isopleth (such as the 0.95 and 0.50 ones classically used to determine the overall distribution and its core areas), the cells of the virtual grid involved in the computation have to be sorted in the decreasing order of the densities and the corresponding fractions has to be summed up until the cumulative isopleth value is reached. The portion of area one looks for then corresponds to the set of the cells involved by this summation. When cells have all the same area, fractions and densities are proportional to each other. Sorting and summing can then be both performed using a single variable (density or fraction). When the cell area is variable, the procedure is just a bit more complex: sorting densities and summing the corresponding fractions must be made jointly in parallel rather than through a single operation.

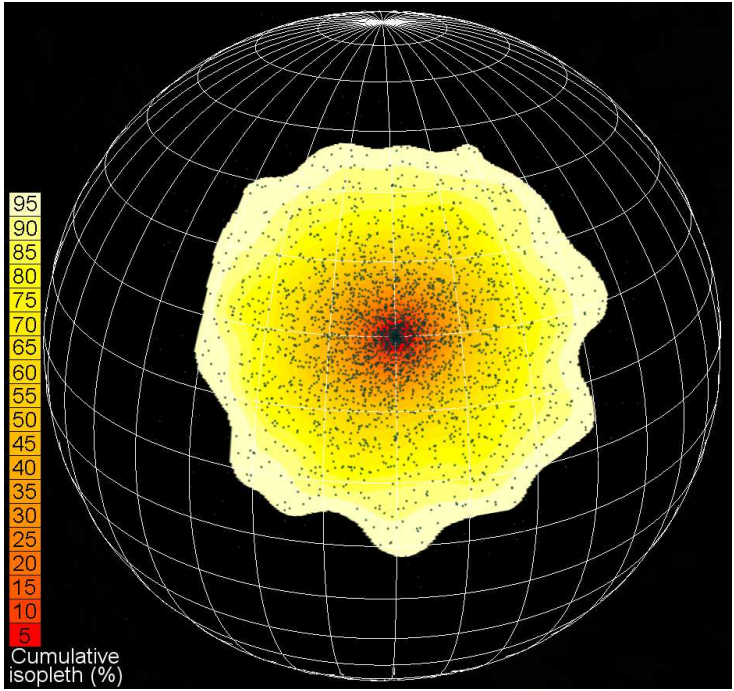

Fig. 5. KDE-based distribution of 3D orientations (i.e. locations on a unit sphere). In this example based on computer simulations, $n = 3000$ locations (dark green dots) were drawn independently of each other from a bivariate von Mises-Fisher distribution centered on location $(0, \pi/6)$ with concentration parameter $\kappa = 31.5$ (corresponding to standard-deviation $\sigma \approx \pi/18$). The KDE was obtained using von Mises-Fisher kernels centered on each location with a smoothing parameter $h = \kappa^{-0.5}/n^{1/6} \approx \pi/68$ (replace $\pi$ by 180 to obtain values in degrees). The density level decreases from the red area to the pale yellow area. The overall area has been cut at the 0.95 cumulative isopleth. In this particular example, any cumulative isopleth is theoretically a circle, but the low density in the peripheral areas makes them highly sensitive to random sampling fluctuations.

Thus, KDE applies to planar and spherical distributions quite similarly. This holds true for both the location-based (i.e. classical) version (LKDE) and the movement-based version (MKDE). MKDE is more suitable than LKDE when locations are highly serially correlated. It extends LKDE by setting kernels (with a variable smoothing parameter) on in-between interpolated locations to simulate biased random bridges linking successive recorded locations (Benhamou 2011). Furthermore, by weighting MKDE by either the local residence time or the local number of visits, the overall distribution can be decomposed in terms of spatial distributions of the frequency of visits and of the mean time spent per visit (Benhamou & Riotte-Lambert 2012). This approach applies as well to the time series of headings or postures considered as a movement on the unit sphere. In this context, the 'number of visits' within an arbitrary angular distance of any given heading or posture $\mu_i$ indicates how often the heading or posture comes back close to $\mu_i$, while the 'residence time' indicates how long it remains close to $\mu_i$ at each (re)visit (see Appendix C). A program designed to compute these different distributions, BRB/MKDE_ORI3D (including LKDE as a particular case), can be freely downloaded from www.cefe.cnrs.fr/fr/recherche/bc/dpb/216-simon-benhamou.

## 8. Fine-scale movement reconstruction

The data acquired by a high-frequency tri-axial accelero-magnetometer, and possibly a speedometer, makes it possible to infer fine-scale movements between successive GNSS relocations of animals that cannot be tracked with a very high frequency. This is in particular the case of marine mammals, sea turtles or penguins, which can be relocated only when they come to the surface to breath. For non-aquatic animals tracked over long periods, a relatively low relocation frequency (e.g. every 30 min) may also have been deliberately chosen to spare the battery of the on-board GNSS receptor. Fine-scale movement reconstruction is often referred to as dead-reckoning (Wilson & Wilson 1988), as it is somewhat similar to the way ancient navigators kept track at sea of the compass-bearing of the harbor.

The successive animal's locations are computed recurrently as $\mathbf{x}_i = (x_i, y_i, z_i) = \mathbf{x}_{i-1} + \mathbf{v}_i \Delta t$, where $\mathbf{v}_i$ is the animal's velocity for the $i^{\text{th}}$ elementary time step $\Delta t$. The norm of this vector corresponds to the speed, $v_i$, and its orientation corresponds to the heading $(\theta_i, \phi_i)$. Such a recurrent procedure is however prone to accumulate random errors due to the various noises affecting the speed, magnetic and acceleration measures, from which $\mathbf{v}_i$ is derived. It can therefore work reliably only for relatively short delays (see Gunner et al. 2021) between ground-truth, GNSS-based relocations used as 'anchors' to compel the reconstructed movement bouts to start and end at these relocations.

The principle for estimating the heading is detailed in §3. However, here heading $(\theta_i, \phi_i)$ applies to the step $\mathbf{x}_{i-1} \rightarrow \mathbf{x}_i$ rather than to the $t_i$ time-stamped location $\mathbf{x}_i$: it should be assessed from the accelero-magnetic data recorded between times $t_{i-1}$ and $t_i$ rather than between $t_i - \Delta t/2$ and $t_i + \Delta t/2$. The forward speed $v_i$ can theoretically be estimated by time-integrating the surge acceleration signal, but this signal is usually far too noisy for this purpose. Despite swimming animals move in a smoother way than terrestrial ones, other methods had to be developed to estimate their speed (with various limitations), e.g. a bending paddle (R. Wilson et al. 2008), a passive propeller (Shiomi et *al*. 2008) or recording the sound level of the water flow against the body (Le Bras et *al*. 2017). For terrestrial animals, a possibility would be to use the visual flow from the ground, but the amount of energy it requires would prevent this technique from being usable continuously for extended periods. Fortunately, as the reconstructed movement will be stretched or shrunk to start and end at GNSS-based 'anchors' (see below), one only needs a proxy $\tilde{v}_i$ that is proportional to the actual speed $v_i$. Without a device that can act as a speedometer, DBA can be considered a speed proxy, with some limitations. Indeed, for numerous species, DBA appears to be roughly proportional to the speed (Bidder et al. 2012). However, passive gliding in air or water is characterized by the same very low DBA as resting, while a high DBA value can be associated with a null speed when the animal is shaking its body on the spot. A detailed analysis of the acceleration and magnetic signals might help highlight these active phases involving a null speed.

Movement reconstruction involves merging the velocity data with ground-truth (GNSS-based) relocations used as 'anchors' to specify the start and end of each reconstructed movement bout. This procedure removes the systematic errors that may affect the azimuth (e.g. due to neglecting the magnetic declination) and the speed (due to considering only a speed proxy proportional to the actual speed). The accelero-magnetometer (and speedometer, if any) measures are usually time-stamped with an independent clock. If so, these measures must be synchronized with GNSS relocations manually, to the closest second, e.g. by inducing a short and strong acceleration at a precise GNSS time, at the beginning of the tracking period, and if possible (battery still alive), at its end, so as to correct for a possible a time drift of the accelero-magnetometer clock. For marine animals breathing at the surface and equipped with a time-depth recorder (TDR) sharing the same clock as the accelero-magnetometer, the synchronization with GNSS locations can also be done based on surfacing events, as no relocation can be recorded when the animal is underwater. I will first detail the general approach that applies to 3D movements. The simplified version that applies to 2D movements of terrestrial animals for which the logger is mounted on a collar rather than tightly mounted on the animal's back will presented afterwards.

Consider the overall movement defined by a series of $N+1$ GNSS-based relocations, expressed in terms of longitudes ($\Xi_I$), latitudes ($\Psi_I$), and altitudes or depths ($Z_I$), recorded at times $T_I$, for $I = 0 \ldots N$. For animals that move over limited extents, longitudes and latitudes can be converted into Cartesian coordinates, ($X_I$, $Y_I$), through a small-extent metric projection (e.g. UTM). Otherwise, the arc-step linking any two successive relocations ($\Xi_{I-1}$, $\Psi_{I-1}$) and ($\Xi_I$, $\Psi_I$) is usually sufficiently short to be suitably converted into a vector-step through a simple local flat-Earth approximation. In this case, a convenient solution for reconstructing the movement bout for the time interval $\Delta T_I = T_I - T_{I-1}$ consists in (i) setting $X_{I-1} = Y_{I-1} = 0$, $X_I = R(\Xi_I - \Xi_{I-1})\cos((\Psi_{I-1} + \Psi_I)/2)$ and $Y_I = R(\Psi_I - \Psi_{I-1})$, where $R = 6,371,000$ m is the mean Earth radius and longitudes and latitudes are expressed in radians, (ii) computing the dead-reckoned Cartesian coordinates $x_i$ and $y_i$ of in-between locations (Eqs **17** or **19**), and (iii) back-converting $x_i$ into longitude $\xi_i = \Xi_{I-1} + x_i/(R\cos((\Psi_{I-1} + \Psi_I)/2))$, and $y_i$ into latitude $\psi_i = \Psi_{I-1} + y_i/R$.

In any case, the $I$th GNSS-based step ($\Delta X_I$, $\Delta Y_I$, $\Delta Z_I$), with $\Delta X_I = X_I - X_{I-1}$, $\Delta Y_I = Y_I - Y_{I-1}$, and $\Delta Z_I = Z_I - Z_{I-1}$, can be represented by the ground-truth vector $L_I[\Theta_I \bullet \Phi_I]$, with length $L_I = ((\Delta X_I)^2 + (\Delta Y_I)^2 + (\Delta Z_I)^2)^{0.5}$, azimuth $\Theta_I = \text{atan}_2(\Delta Y_I, \Delta X_I)$ and elevation angle $\Phi_I = \sin^{-1}(\Delta Z_I/L_I)$. This step can also be assessed as the dead-reckoned vector $\tilde{V}_I \Delta t [\widehat{\Theta}_I \bullet \widehat{\Phi}_I] = \Sigma_i \tilde{v}_i \Delta t [\theta_i \bullet \phi_i]$, for $i = 1 \ldots n_I = \text{round}(\Delta T_I/\Delta t)$, where $\tilde{v}_i$, $\theta_i$ and $\phi_i$ are the speed proxy, body azimuth and body elevation angle estimated for the $i$th elementary time step $\Delta t$: $\tilde{V}_I = (\Sigma_i^2 \tilde{v}_{xi} + \Sigma_i^2 \tilde{v}_{yi} + \Sigma_i^2 \tilde{v}_{zi})^{0.5}$, $\widehat{\Theta}_I = \text{atan}_2(\Sigma_i \tilde{v}_{yi}, \Sigma_i \tilde{v}_{xi})$ and $\widehat{\Phi}_I = \sin^{-1}(\Sigma_i \tilde{v}_{zi}/\tilde{V}_I)$, with $\tilde{v}_{xi} = \tilde{v}_i \cos(\phi_i) \cos(\theta_i)$, $\tilde{v}_{yi} = \tilde{v}_i \cos(\phi_i) \sin(\theta_i)$ and $\tilde{v}_{zi} = \tilde{v}_i \sin(\phi_i)$. Systematic errors and the accumulation of random errors result in a dead-reckoned vector that does not match the ground-truth vector. As all these errors come from misestimates in terms of orientation and length, the dead-reckoned movement must be corrected in polar (rather than Cartesian, as proposed by R. Wilson et al. 2008) terms, through rotation and rescaling (i.e. stretching or shrinking), to preserve its overall shape.

The orientation error can be corrected by rotating the dead-reckoned vector first by $-\widehat{\Theta}_I$ around the $X$ axis (i.e. in the horizontal plane) to get its azimuth pointing eastwards, then by $\Phi_I - \widehat{\Phi}_I$ around the $Y$ axis (i.e. in the E-W vertical plane) to correct its elevation angle without changing its azimuth, and finally by $\Theta_I$ around the $X$ axis (i.e. in the horizontal plane) to fully align it onto the ground truth vector. The error in length can be corrected by rescaling the dead-reckoned vector by $L_I/(\tilde{V}_I\Delta t)$. The orientation adjustment and rescaling apply similarly to every step. With $\mathbf{x}_0$ set to $\mathbf{X}_{I-1}$, the series of subsequent locations $\mathbf{x}_i = (x_i, y_i, z_i)$, for $i = 1 \ldots n_I$, specifying the path performed between the two successive ground-truth relocations, can therefore be computed recurrently as $\mathbf{x}_i^\top = \mathbf{x}_{i-1}^\top + \mathbf{R}_3(\Theta_I)\mathbf{R}_2(\Phi_I - \widehat{\Phi}_I)\mathbf{R}_3(-\widehat{\Theta}_I)\tilde{\mathbf{v}}_i^\top L_I/\tilde{V}_I$, i.e.

$$x_i = x_{i-1} + (r_i \cos(\Theta_I) - q_i \sin(\Theta_I))\, L_I/\tilde{V}_I \tag{17a}$$

$$y_i = y_{i-1} + (q_i \cos(\Theta_I) + r_i \sin(\Theta_I))\, L_I/\tilde{V}_I \tag{17b}$$

$$z_i = z_{i-1} + (\tilde{v}_{zi} \cos(\Phi_I - \widehat{\Phi}_I) + p_i \sin(\Phi_I - \widehat{\Phi}_I))\, L_I/\tilde{V}_I \tag{17c}$$

with $p_i = \tilde{v}_{xi} \cos(\widehat{\Theta}_I) + \tilde{v}_{yi} \sin(\widehat{\Theta}_I)$, $q_i = \tilde{v}_{yi} \cos(\widehat{\Theta}_I) - \tilde{v}_{xi} \sin(\widehat{\Theta}_I)$ and $r_i = p_i \cos(\Phi_I - \widehat{\Phi}_I) - \tilde{v}_{zi} \sin(\Phi_I - \widehat{\Phi}_I)$. This formulation warrants that the dead-reckoned movement between times $T_{I-1}$ and $T_I$ starts at $\mathbf{X}_{I-1} = (X_{I-1}, Y_{I-1}, Z_{I-1})$ and ends at $\mathbf{X}_I = (X_I, Y_I, Z_I)$ The location errors at intermediate times tend to increase with the time interval $\Delta T_I$, which should therefore be kept as small as possible, in particular when the speed is poorly estimated.

For marine animals that breathe at the sea surface, Eqs (**17**) apply in a simplified form with $Z_{I-1} = Z_I = 0$ and therefore $\Phi_I = 0$ and $L_I = ((\Delta X_I)^2 + (\Delta Y_I)^2)^{0.5}$. However, it is worth noting that $\tilde{v}_i$, $\theta_i$ and $\phi_i$ concern the swimming (i.e. water masses-related) movement, whereas the GNSS-based relocations used as 'anchors', $\mathbf{X}_{I-1}$ and $\mathbf{X}_I$, are expressed in a ground-bound frame of reference. Consequently, if the current speed is not negligible with respect to the animal's swimming speed, Eqs (**17**) apply with $L_I = ((\Delta X_I - \Delta T_I E_I)^2 + (\Delta Y_I - \Delta T_I N_I)^2)^{0.5}$ and $\Theta_I = \mathrm{atan}_2(\Delta Y_I - \Delta T_I N_I, \Delta X_I - \Delta T_I E_I)$, where $E_I$ and $N_I$ are the mean eastward (zonal) and northward (meridional) components of the current velocity experienced by the animal for the time interval $\Delta T_I$, so as to subtract the deflecting effect of the current (see details in Girard et al. 2006). The time series of dead-reckoned $x_i$ and $y_i$ values (Eqs **17**a and **17**b) then specify the swimming movement in the horizontal plane. The corresponding geographical coordinates are $x_i^g = x_i + i\Delta t E_I$ and $y_i^g = y_i + i\Delta t N_I$. Furthermore, the dead-reckoned $z_i$ values (Eq. **17**c) should be replaced by TDR-based ones, if available, as they are more accurate.

When the device is mounted on the collar of a terrestrial animal, the device's posture can dramatically differ from the body posture, and vary during movement because the collar is liable to roll around the neck. However, mounting the device with its surge axis orthogonal to the collar plane warrants that this axis remains in (or parallel to) the animal's sagittal plane whatever the collar orientation. Even so, the body azimuth will differ from the device's azimuth whenever the animal moves banked (i.e. $\lambda \neq 0$) if the body and device's surge axes do not perfectly match. Fortunately, terrestrial animals can be expected to always move unbanked ($\lambda \approx 0$), so that the body azimuth corresponds to the device's azimuth, and can therefore be computed, similarly to Eq. (**5**), as

$$\theta = \mathrm{atan}_2((\overline{a}_V^2 + \overline{a}_W^2)\overline{m}_U - \overline{a}_U(\overline{a}_V\overline{m}_V + \overline{a}_W\overline{m}_W),\ (\overline{a}_V\overline{m}_W - \overline{a}_W\overline{m}_V)\|\overline{\mathbf{a}}\|). \tag{18}$$

The $x$ and $y$ coordinates on the horizontal plane are then obtained recurrently by applying Eqs (**17**) with $\Phi_I = \widehat{\Phi}_I = \Delta Z_I = 0$, and $\tilde{v}_{zi} = 0$ for any $i$, i.e. simply as:

$$x_i = x_{i-1} + (\cos(\Theta_I - \widehat{\Theta}_I)\,\tilde{v}_{xi} - \sin(\Theta_I - \widehat{\Theta}_I)\,\tilde{v}_{yi})\, L_I/\tilde{V}_I \tag{19a}$$

$$= x_{i-1} + ((\Delta X_I \Sigma_i \tilde{v}_{xi} + \Delta Y_I \Sigma_i \tilde{v}_{yi})\,\tilde{v}_{xi} - (\Delta Y_I \Sigma_i \tilde{v}_{xi} - \Delta X_I \Sigma_i \tilde{v}_{yi})\,\tilde{v}_{yi})/(\Sigma_i^2 \tilde{v}_{xi} + \Sigma_i^2 \tilde{v}_{yi})$$

$$y_i = y_{i-1} + (\cos(\Theta_I - \widehat{\Theta}_I) \ \tilde{v}_{yi} + \sin(\Theta_I - \widehat{\Theta}_I) \ \tilde{v}_{xi}) \ L_I/\tilde{V}_I \qquad (\mathbf{19}b)$$

$$= y_{i-1} + ((\Delta X_I \Sigma_i \tilde{v}_{xi} + \Delta Y_I \Sigma_i \tilde{v}_{yi}) \ \tilde{v}_{yi} + (\Delta Y_I \Sigma_i \tilde{v}_{xi} - \Delta X_I \Sigma_i \tilde{v}_{yi}) \ \tilde{v}_{xi})/(\Sigma_i^2 \tilde{v}_{xi} + \Sigma_i^2 \tilde{v}_{yi}).$$

with $\tilde{v}_{xi} = \tilde{v}_i\cos(\theta_i)$, $\tilde{v}_{yi} = \tilde{v}_i\sin(\theta_i)$, $L_I = ((\Delta X_I)^2 + (\Delta Y_I)^2)^{0.5}$, and $\tilde{V}_I = (\Sigma_i^2 \tilde{v}_{xi} + \Sigma_i^2 \tilde{v}_{yi})^{0.5}$. An example is shown in Fig. 6. If the $z$ coordinates are relevant (mountainous environment), they can be inferred from the $(x,y)$ coordinates and a Digital Elevation Model. When estimated at a relatively high frequency (e.g. with $\Delta t = 1$ s) the two speed components, $\tilde{v}_{xi}$ and $\tilde{v}_{yi}$, tends to be very noisy because the rough locomotion mode of terrestrial animals tends to strongly shake the device. Some smoothing of the path is therefore often necessary.

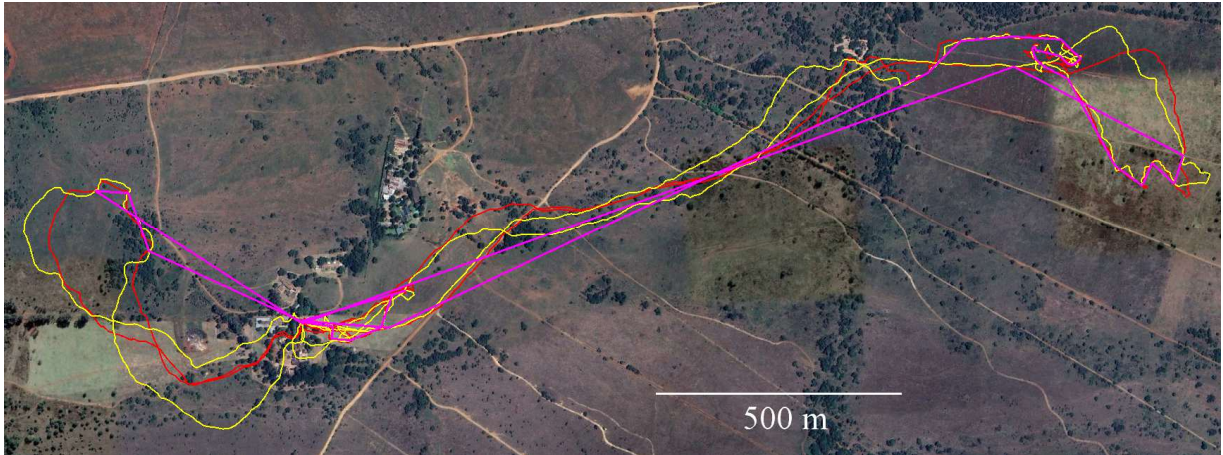

**Fig. 6. Reconstructed path of a domestic free-ranging horse** (based on unpublished data recorded by Jodie Martin). It was GPS-tracked for 24 h with a relocation every 10 s, to obtain a faithful path representation (in red). The GPS relocations were subsampled to one every 30 min. to obtain the kind of path representation that is usually obtained when tracking wild animals (in magenta). Then, the data acquired by an on-board accelero-magnetometer were used to reconstruct the path between 30-min GPS relocations (in yellow) using Eqs (**19**).

## 9. Conclusions

On-board tri-axial high-frequency accelerometers, magnetometers and gyrometers, alone or coupled, have become cheaper and less energy-consuming. This makes it possible to record data on more individuals belonging to smaller species or for longer durations. The present was written with the aim to help the researchers who wish to address questions that require this type of data to compute the key variables and represent their distributions more easily.

Beyond the various trigonometric expressions that are required to compute heading/posture (and possibly to relate their changes in terms of body rotations) from raw accelero-magnetic data, a recurrent issue with tri-axial high-frequency devices working continuously comes from the huge amount of raw data to manage. It is however worth noting that the successive raw data values acquired at high-frequency are both noisy and highly serially correlated. This means that amount of information they provide is much lower than the amount of data. The derived variables such as heading and posture can be estimated at the same high frequency $f$ as the one used to acquire the raw data using a $\Delta t$-width *sliding* window which can dramatically smooths the signal (i.e. remove a lot of noise). However, this procedure also dramatically increases the level of serial correlation mechanically (with a sliding window encompassing $k = \text{round}(f\Delta t)$ values, the range of values considered a given time and at the next time shared $k$–1 values), so that the information provided by the change of heading and posture for the very short time $1/f$ is likely to be insignificant.

Using a $\Delta t$-width *jumping* rather than *sliding* window will smooth the signal as well. The amount of data to store and to process in subsequent analyses is then dramatically reduced (by $f\Delta t$, a single value of heading/posture being kept for each time step $\Delta t$), and the level of serial correlation is reduced as well, so that most information is preserved. In this case, the time step $\Delta t$ should be not only large enough to flatten body oscillations due to limb movements (Shepard et al. 2008a), but also short enough to preserve the highest possible resolution and warrants that (i) changes of heading/posture can be faithfully represented by orthodromic arcs, (ii) roll remains moderate ($|\rho|<\pi$) and (iii) reorientation and rolling speeds remain approximately constant (but they can markedly differ between time steps). In-between posture/heading can no longer be directly estimated, but they can be interpolated, if required, using Eqs (**6-8**). Furthermore, Eqs. (**11-13**) highlight how body rotations and changes of posture/heading are interrelated for the time step $\Delta t$ considered. For technical reasons, posture must be initially computed at high-frequency using a *sliding* window to assess DBA (at the same high frequency). However, only a single value of posture (through sub-sampling) and a single value of DBA (through averaging) need to be kept for each time step $\Delta t$. It is also worth noting that the activity level can be reliably estimated using a low-frequency accelerometer (Benoit et al. 2020). A high-frequency accelerometer is therefore required only if one also wishes to reliably assess the posture.

It may be necessary to keep the whole time series of raw data for performing complex analyses such as spectral analyses to detect and characterize regular body oscillations (e.g. Williams et al. 2017). However, there often exist simpler and lighter alternatives. For instance, such oscillations in a flying bird, due to wing movements, can be detected and characterized more simply by keeping only two values for each time step $\Delta t$: the mean $\bar{d}$ and the coefficient of variation, CV, of the delay between successive times at which the heave component of acceleration goes through a balance point equal to $-g$. Random fluctuations involves that the delay is exponentially distributed (CV=1). Regular oscillations can therefore be detected by CV<<1, and their period can then be estimated as $T = 2\bar{d}$ (for $\bar{d}>1/f$; Nyquist criterion). Based on this approach and on the value of the elevation angle $\phi$, kestrel falcons could be said to hover (CV<0.5, $T$<150 ms), perform flapped ($\phi \approx 0$, CV<0.5, $T\in$[150, 220] ms) or glided ($\phi \approx 0$, CV>0.5) fly, or to be perched ($\phi$>>0, CV>0.5; Daniel García Silveira, unpublished results). For sea turtles, which swim using a forward-backward movement of front flippers, a similar analysis can be done by considering the surge component of acceleration with respect to a balance point equal to $-\sin(\phi)g$.

As these examples show, some behaviors can be inferred by comparing values derived from the data acquired by bio-loggers to relevant thresholds (see also Shepard et al. 2008b, R. Wilson et al. 2008, Fluhr et al. 2021). However, when quite different behaviors generate relatively similar signals, this simple approach does not work. Machine learning then can help infer behavior from 'big data' acquired by on-board loggers (Jeantet et al. 2020, 2021). It should nevertheless be keep in mind that learning will be faster and/or more easily applicable if the machine is fed with biologically relevant parameters such as, for example, DBA and elevation and bank angles rather than with raw acceleration data. Correcting the parameters that apply to the device to obtain those that apply to the body (Eqs **1**) should also avoid that the results obtained were blurred by the discrepancy between the two frames of references, which may vary between individuals. In other terms, an approach based on artificial intelligence will work all the better when supported by human intelligence.

## REFERENCES

Benhamou S. 2011. Dynamic approach to space and habitat use based on biased random bridges. *PLoS ONE* 6, e14592.

Benhamou S. 2014. Of scales and stationarity in animal movements. *Ecol. Let.* 17, 261–272.

Benhamou S. 2018. Mean squared displacement and sinuosity of three-dimensional random search movements. *arXiv*:1801.02435

Benhamou S. & Riotte-Lambert L. 2012. Beyond the utilization distribution: Identifying home range areas that are intensively exploited or repeatedly visited. *Ecol. Model.* 227: 112–116.

Benoit L. et al. 2020. Accelerating across the landscape: the energetic costs of natal dispersal in a large herbivore. *J. Anim. Ecol.* 89: 173–185.

Bidder O.R. et al. 2012. The need for speed: testing acceleration for estimating animal travel rates in terrestrial dead-reckoning systems. *Zool.* 115: 58–64

Fisher R.A 1953. Dispersion on a sphere. *Proc. R. Soc. Lond.* A 217: 295–305.

Fluhr J., Benhamou S., Peyrusqué D. & Duriez O. 2021. Space use and time budget in two populations of griffon vultures in contrasting landscapes. *J. Raptor Res.* 55: 425–437.

Girard C., Sudre J., Benhamou S., Roos D & Luschi P. 2006. Homing in green turtles (*Chelonia mydas*): oceanic currents act as a constraint rather than as an information source. *Mar. Ecol. Prog. Ser.* 322: 281–289.

Gleiss A.C., Wilson R.P. & Shepard E.L. 2011. Making overall dynamic body acceleration work: on the theory of acceleration as a proxy for energy expenditure. *Meth. Ecol. Evol.* 2: 23–33.

Gunner R.M. et al. 2021. How often should dead-reckoned animal movement paths be corrected for drift? *Anim. Biotelem.* 9:43

Jeantet L. et al. 2020. Behavioural inference from signal processing using animal-borne multi-sensor loggers: a novel solution to extend the knowledge of sea turtle ecology. *R. Soc. Open Sci.*7: 200139.

Jeantet L., Vigon V., Geiger S. & Chevallier D. 2021. Fully convolutional neural network: A solution to infer animal behaviours from multi-sensor data. *Ecol. Model.* 450: 109555.

Johnson M.P. & Tyack P.L 2003. A digital acoustic recording tag for measuring the response of wild marine mammals to sound. *IEEE J. Ocean. Engin*. 28: 3–12.

Le Bras Y., Jouma'a J. & Guinet C. 2017. Three-dimensional space use during the bottom phase of southern elephant seal dives. *Mov. Ecol.* 5: 18.

Qasem L. et al. 2012. Tri-axial dynamic acceleration as a proxy for animal energy expenditure; should we be summing values or calculating the vector? *PLoS ONE* 7: e31187.

Shepard E.L. et al. 2008a. Derivation of body motion via appropriate smoothing of acceleration data. *Aquat. Biol.* 4: 235–241.

Shepard E.L. et al. 2008b. Identification of animal movement patterns using tri-axial accelerometry. *Endang. Species Res.* 10: 47–60.

Shiomi K. et al. 2008. Effect of ocean current on the dead-reckoning estimation of 3-D dive paths of emperor penguins. *Aquat. Biol.* 3: 265–270.

Silverman B.W. 1986. *Density estimation for statistics and data analysis*. Boca Raton, Florida, USA: Chapman & Hall/CRC. 175 p.

Williams H.J., Shepard E.L., Duriez O. & Lambertucci S.A. 2015. Can accelerometry be used to distinguish between flight types in soaring birds? *Anim. Biotelem.* 3: 45.

Williams H.J. et al. 2017. Identification of animal movement patterns using tri-axial magnetometry. *Mov. Ecol.* 5: 6.

Williams H.J. et al. 2018. Vultures respond to challenges of near-ground thermal soaring by varying bank angle. *J. Exp. Biol.* 221: jeb174995.

Williams H.J. et al. 2020. Optimising the use of bio-loggers for movement ecology research. *J. Anim. Ecol.* 89: 186–206.

Wilson A.M. et al. 2013. Locomotion dynamics of hunting in wild cheetahs. *Nature* 498: 185–189.

Wilson R.P., Shepard E.L. & Liebsch N. 2008. Prying into the intimate details of animal lives: use of a daily diary on animals. *Endang. Species Res*. 4: 123–137.

Wilson, R.P. & Wilson, M.-P. 1988. Dead reckoning: a new technique for determining penguin movements at sea. *Meeresforschung – Rep. Mar. Res.* 32: 155–158.

Wilson R.P. et al. 2006. Moving towards acceleration for estimates of activity-specific metabolic rate in free-living animals: the case of the cormorant. *J. Anim. Ecol.* 75: 1081–1090.

Wilson R.P. et al. 2007. All at sea with animal tracks; methodological and analytical solutions for the resolution of movement. *Deep-Sea Res.* II 54: 193–210.

Wilson R.P. et al. 2016. A spherical-plot solution to linking acceleration metrics with animal performance, state, behaviour and lifestyle. *Mov. Ecol*. 4: 22.

Wilson R.P. et al. 2020a. Estimates for energy expenditure in free-living animals using acceleration proxies: A reappraisal. *J. Anim. Ecol*. 89:161–172.

Wilson R.P. et al. 2020b. An "orientation sphere" visualization for examining animal head movements. *Ecol. Evol*. 10:4291–4302.

# APPENDICES

## A) Deriving the body-related vector from an animal-borne device-related vector

The vector $\mathbf{D} = (D_U, D_V, D_W)$ that applies to an animal-borne device corresponds to the vector $\mathbf{B} = (B_U, B_V, B_W)$ that applies to the free moving animal only if the surge, sway and heave axes of the animal's body and of the device match each other. The device may have been set tilted longitudinally by $\phi^*$ and/or transversally by $\lambda^*$. Furthermore, the device's azimuth at this time, $\theta^*$, may also differ from the actual, *visually assessed* body azimuth, $\theta^{obs}$, if the device was not set in such a way its surge axis lies in the body's sagittal plane. In the general case, $\mathbf{B}$ can be derived from $\mathbf{D}$ through the relationship $\mathbf{B}^T = \mathbf{R}_3(\theta^* - \pi/2 + \theta^{obs})\mathbf{R}_2(\phi^*)\mathbf{R}_1(\lambda^*)\mathbf{D}^T$, based on the acceleration and calibrated magnetic vectors, $\mathbf{a}^* = (a_U^*, a_V^*, a_W^*)$ and $\mathbf{m}^* = (m_U^*, m_V^*, m_W^*)$, recorded by the device while the animal rests level both longitudinally and transversally ($\phi \approx \lambda \approx 0$, $\phi^* = \sin^{-1}(-a_U^*/\|\mathbf{a}^*\|)$, $\lambda^* = \mathrm{atan}_2(-a_V^*, -a_W^*)$, $\theta^* = \mathrm{atan}_2((a_V^{*2} + a_W^{*2})m_U^* - a_U^*(a_V^* m_V^* + a_W^* m_W^*)$, $(a_V^* m_W^* - a_W^* m_V^*)\|\mathbf{a}^*\|))$, with body azimuth $\theta^{obs}$ assumed to be visually assessed using a magnetic compass, i.e. measured positively clockwise from North (the corresponding value measured positively counterclockwise from East, as $\theta^*$, is therefore $\pi/2 - \theta^{obs}$). Determining visually when an animal rests level is relatively easy, but visually assessing its magnetic azimuth $\theta^{obs}$ at this time can be tricky. Setting, whenever possible, the device in such a way its surge axis lies in the body sagittal plane warrants that the device's azimuth corresponds to the body azimuth ($\theta^* \approx \pi/2 - \theta^{obs}$) when the animal is level ($\phi \approx \lambda \approx 0$), even if the device was tilted longitudinally ($\phi^* \neq 0$) and/or transversally ($\lambda^* \neq 0$). There is then no more need to visually assess $\theta^{obs}$ nor to compute $\phi^*$: $\mathbf{B}^T = \mathbf{R}_2(\phi^*)\mathbf{R}_1(\lambda^*)\mathbf{D}^T$ (Eqs **1**). In any case, the conversion of $\mathbf{D}$ into $\mathbf{B}$ can apply to an accelerometer (with $\mathbf{B}$ replaced by $\mathbf{A}$ and $\mathbf{D}$ by $\bar{\mathbf{a}}$), a calibrated magnetometer (with $\mathbf{B}$ replaced by $\mathbf{M}$ and $\mathbf{D}$ by $\bar{\mathbf{m}}$) or a gyrometer (with $B_U$ replaced by $\dot{\rho}$, $B_V$ by $-\dot{\beta}$, $B_W$ by $\dot{\alpha}$, $D_U$ by $\dot{\rho}^d$, $D_V$ by $-\dot{\beta}^d$ and $D_W$ by $\dot{\alpha}^d$). For collared terrestrial animals, which are expected to always move unbanked ($\lambda \approx 0$), mounting the device with its surge axis orthogonal to the collar plane at least warrants that the device azimuth reflects the body azimuth, which can therefore be assessed directly from $\bar{\mathbf{a}}$ and $\bar{\mathbf{m}}$ (Eq. **18**), even though the body posture can be hardly assessed.

## B) Expressing the 3D orientation of the head relatively to the body

Setting a high-frequency tri-axial accelero-magnetometer on the back and another on the head makes it possible to compute not only the 3D orientations in absolute space of both the body ($\theta_b$, $\phi_b$, $\lambda_b$) and the head ($\theta_h$, $\phi_h$, $\lambda_h$), independently of each other (Eqs **3** and **5**), but also the body-related 3D orientation of the head ($\theta_{h/b}$, $\phi_{h/b}$, $\lambda_{h/b}$). For this purpose, the surge, sway and heave axes of the head are represented in absolute space by the units vectors $\mathbf{x}_h = [\theta_h \bullet \phi_h]$, $\mathbf{x}'_h = [\theta'_h \bullet \phi'_h]$ and $\mathbf{x}''_h = [\theta''_h \bullet \phi''_h]$. The Cartesian coordinates of $\mathbf{x}_h$ and $\mathbf{x}'_h$, derived from the relationships $\mathbf{x}_h^T = \mathbf{R}_3(\theta_h)\mathbf{R}_2(\phi_h)\,(1, 0, 0)^T$ and $\mathbf{x}_h'^T = \mathbf{R}_3(\theta_h)\mathbf{R}_2(\phi_h)\mathbf{R}_1(\lambda_h)\,(0, 1, 0)^T$, can be re-expressed in the body-bound frame of reference using the relationships $\mathbf{x}_{h/b}^T = \mathbf{R}_1(-\lambda_b)\mathbf{R}_2(-\phi_b)\mathbf{R}_3(-\theta_b)\mathbf{x}_h^T$

and $\mathbf{x}'_{h/b}{}^T = \mathbf{R}_1(-\lambda_b)\mathbf{R}_2(-\phi_b)\mathbf{R}_3(-\theta_b)\mathbf{x}'_h{}^T$. The body-related 3D orientation of the head is then specified by azimuth $\theta_{h/b} = \text{atan}_2(y_{h/b}, x_{h/b})$, elevation angle $\phi_{h/b} = \sin^{-1}(z_{h/b})$, and bank angle $\lambda_{h/b} = \text{atan}_2(z'_{h/b}, z''_{h/b})$, with $z''_{h/b} = x_{h/b}\, y'_{h/b} - y_{h/b}\, x'_{h/b}$ (based on $\mathbf{x}'' = \mathbf{x} \wedge \mathbf{x}'$, where $\wedge$ stands for the cross product). In the particular case of an animal that keeps its head (almost) orthogonal to the body coronal plane (i.e. $\phi_{h/b} \approx \pm\pi/2$, $x_{h/b} \approx y_{h/b} \approx z'_{h/b} \approx z''_{h/b} \approx 0$), the functions $\text{atan}_2(y_{h/b}, x_{h/b})$ and $\text{atan}_2(z'_{h/b}, z''_{h/b})$ will result in irrelevant values, but the body-related 3D orientation of the head can be consistently specified by 'azimuth' $\theta_{h/b} = \theta'_{h/b} - \pi/2$, with $\theta'_{h/b} = \text{atan}_2(y'_{h/b}, x'_{h/b})$, elevation angle $\phi_{h/b} = \text{sgn}(z_{h/b})\,\pi/2$, and bank angle $\lambda_{h/b} = 0$. The above relationships can also be used, for example, to determine how a given individual is perceived by another individual (sexual partner, competitor, predator or prey) by computing the 3D orientation of the body of the former relatively to the head of the latter.

### C) 'Residence Time' and 'Number of Visits' for Heading or Posture

The residence time and the number of visits attributed to a given 'location' (heading or posture) $\boldsymbol{\mu}_i$ on the unit sphere are computed, as for a location on the plane, from the passage times when the 'movement' enters and exits a virtual $\boldsymbol{\mu}_i$-centered circle with an arbitrary ad hoc radius. The residence time, $\text{RT}(i)$, is the sum of the first crossing duration, $F_1(i) - B_1(i)$, where $F_1(i)$ and $B_1(i)$ are the first forward and backward passage times, and (to get a steadier signal; Barraquand & Benhamou 2008, *Ecol.* 89: 3336–3348), of subsequent crossings of the circle, in backward and forward directions, provided they occurred before the time spent out the circle is larger than a given ad hoc threshold $T_{out}$:

$$\text{RT}(i) = F_1(i) - B_1(i)\ +\ \sum_{j=1}^{Max_F(i)}\Big[F_{2j+1}(i) - F_{2j}(i)\Big] + \sum_{j=1}^{Max_B(i)}\Big[B_{2j}(i) - B_{2j+1}(i)\Big]$$

where $F_j(i)$ and $B_j(i)$ are the forward and backward, respectively, $j^{th}$ passage times through the circle, and $Max_F(i)$ and $Max_B(i)$ indicate that, independently of each other, the forward and backward summations start only if, and continue only while, the time spent outside the circle before re-entering it was smaller than the predefined time threshold (i.e. $F_{2j}(i) - F_{2j-1}(i) < T_{out}$ and $B_{2j-1}(i) - B_{2j}(i) < T_{out}$, respectively). The crossings occurring earlier (backward movements) and later (forward movements) are taken into account in the computation of the residence time associated with other (earlier and later) visits to $\boldsymbol{\mu}_i$. The number of visits to $\boldsymbol{\mu}_i$ is therefore $\text{NV}(i) = 1 + \text{NF}(i) + \text{NB}(i)$, where $\text{NF}(i)$ corresponds to the number of times an even forward passage time, $F_{2j}(i)$, occurs after a delay larger than $T_{out}$ since the preceding odd passage time, $F_{2j-1}(i)$ (i.e. $F_{2j}(i) - F_{2j-1}(i) > T_{out}$), and $\text{NB}(i)$ corresponds to the number of times an odd backward passage time, $B_{2j-1}(i)$, occurs after a delay larger than $T_{out}$ since the preceding even passage time, $B_{2j}(i)$ (i.e. $B_{2j-1}(i) - B_{2j}(i) > T_{out}$). Any passage time, $t_P$, is computed by linear interpolation from the two time-stamped 'locations' recorded just before and just after. For a given passage location $\mathbf{P}$ in the backward (entrance) or forward (exit) direction at an arbitrary ad hoc angular distance $\omega_{P\boldsymbol{\mu}}$ of $\boldsymbol{\mu}_i = (\xi_i, \psi_i)$, let $\mathbf{I} = (\xi_I, \psi_I)$ and $\mathbf{O} = (\xi_O, \psi_O)$, with associated times $t_I$ and $t_O$, be either the last inside and first outside relocations (exit: $t_O > t_I$), or the first inside and last outside locations (entrance: $t_O < t_I$) relocations ($\omega_{I\mu} < \omega_{P\mu} < \omega_{O\mu}$). In both cases, **P** corresponds to the point where the perimeter of the $\mu_i$-centered cap (portion of sphere) with curved radius $\omega_{P\mu}$ is intersected by the arc $\widehat{IO}$, whose size and initial inclination are (similar to Eqs **7**):

$$\omega_{IO} = \cos^{-1}(\sin(\psi_I)\sin(\psi_O) + \cos(\psi_I)\cos(\psi_O)\cos(\xi_O - \xi_I))$$

$$\delta_{IO}^i = \text{atan}_2(\sin(\psi_O) - \cos(\omega_{IO})\sin(\psi_I), \cos(\psi_I)\cos(\psi_O)\sin(\xi_O - \xi_I)).$$

Applying the spherical law of cosines to the triangle $\widehat{\boldsymbol{\mu}I} \cdot \widehat{IP} \cdot \widehat{P\boldsymbol{\mu}}$ leads to the equation $a\cos(\omega_{IP}) + b\sin(\omega_{IP}) = c$ with $a = \cos(\omega_{\mu I})$, $b = \sin(\omega_{\mu I})\cos(\pi - \delta_{IO}^i + \delta_{\mu I}^f)$, and $c = \cos(\omega_{P\mu})$. It has two solutions, which correspond to the two points where the great circle passing through **I** and **O** intersects the small circle representing the perimeter of the $\mu_i$-centered cap. With $\omega_{I\mu} < \omega_{P\mu} < \pi/2$, the single one that corresponds to **P** lying on the arc $\widehat{IO}$ is

$$\omega_{IP} = \tan^{-1}(b/a) + \cos^{-1}(\cos(\omega_{P\mu})/(a^2 + b^2)^{0.5})$$

with $a = \sin(\psi_I)\sin(\psi_i) + \cos(\psi_I)\cos(\psi_i)\cos(\xi_i - \xi_I)$

and $b = \cos(\delta_{IO}^i)\cos(\psi_i)\sin(\xi_i - \xi_I) + \sin(\delta_{IO}^i)(\sin(\psi_i) - a\sin(\psi_I))/\cos(\psi_I)$.

The passage time at **P**, i.e. the entrance time into, or exit time out of, the $\mu_i$-centered cap, is:

$$t_P = t_I + (t_O - t_I)\,\omega_{IP}/\omega_{IO}$$

If necessary, the coordinates of **P** can be computed as (similar to Eqs **6**):

$$\xi_P = \xi_I + \text{atan}_2(\cos(\delta_{IO}^i), \cos(\psi_I)/\tan(\omega_{IP}) - \sin(\delta_{IO}^i)\sin(\psi_I))$$

$$\psi_P = \sin^{-1}(\sin(\omega_{IP})\sin(\delta_{IO}^i)\cos(\psi_I) + \cos(\omega_{IP})\sin(\psi_I)).$$